\RequirePackage[l2tabu, orthodox]{nag}
\RequirePackage{snapshot}

\documentclass[10pt,onecolumn]{extarticle}

\makeatletter
\def\@font@info#1{}
\makeatother

\usepackage[dvips,
            a4paper,
            margin=1in,
            includefoot,
            heightrounded,
            ]{geometry}

\usepackage[english]{babel}

\usepackage{amsmath}
\usepackage{amssymb}
\usepackage{amsfonts}
\usepackage{mathtools}
\usepackage{amsthm}

\usepackage{booktabs}

\usepackage{abstract}

\usepackage{graphicx}
\usepackage[dvipsnames,svgnames,x11names,table]{xcolor}
\usepackage{subfigure}

\usepackage{booktabs}
\usepackage{setspace}

\usepackage{multicol}

\usepackage{cite}
\usepackage{hyperref}

\usepackage[normalem]{ulem}
\usepackage{soul}

\usepackage{enumerate}
\usepackage{multirow}

\usepackage[short,12hr]{datetime}

\usepackage{siunitx}
       \usepackage{fouriernc}

\newcommand{\keywordname}{Keywords}

\newcommand{\printtitle}{%
  \maketitle
  \begin{onecolabstract}
    \myabstract
  \begin{center}
    \small
    \textbf{\keywordname}
    \\\medskip
    \mykeywords
  \end{center}
  \end{onecolabstract}
  \bigskip
  \onehalfspacing
}

\author{%
L.~Andrade-Silva%
\thanks{%
    L.~Andrade-Silva
    is with the Graduate Program in Industrial Engineering,
    Universidade Federal de Pernambuco (UFPE), Caruaru, Brazil,
    and
	the Industrial Signal Processing Laboratory,
	Department of Technology,
	UFPE,
	Caruaru 55014-900, Brazil.
    }
\and
W.~A.~S.~Aleixo%
\thanks{%
    W.~A.~S.~Aleixo
    is with the Graduate Program in Electrical Engineering,
    UFPE, Recife, Brazil,
    and
	the Industrial Signal Processing Laboratory,
	Department of Technology,
	UFPE,
	Caruaru 55014-900, Brazil.
    }
\and
R.~J.~Cintra%
\thanks{%
    R. J. Cintra
    is with
	the Industrial Signal Processing Laboratory,
	Department of Technology,
	UFPE,
	Caruaru 55014-900, Brazil
    (e-mail: rjdsc@de.ufpe.br).}
}

\title{%
32-point DFT Approximations
\\
Based on Minimal Frobenius Error
and
DFT Symmetries}

\newcommand{\myabstract}{%
This work introduces low-complexity, multiplierless approximations for the
32-point discrete Fourier transform.
The proposed methods are obtained
by
minimizing the Frobenius error
compared against the DFT matrix
over a set of trivial multipliers.
A
row-wise, symmetry-constrained parameterization
is employed to reduce the
search space size,
rendering the task computationally
tractable.
The resulting approximations
could outperform
the reference method
in the literature
according to
energy-based error measurements.
A sparse matrix factorization
is provided for efficient computation;
the arithmetic costs
are
152~real additions and 34~bit-shifts
only.
}

\newcommand{\mykeywords}{%
Low-complexity design,
Approximate DFT,
Fast algorithms,
Spectral estimation
}

\date{}

\begin{document}

\printtitle

\section{Introduction}

\subsection{Background}

The $N$-point discrete Fourier transform (DFT) is a linear operator that
maps a discrete-time signal~$x[n]$,
$n = 0, 1, \ldots, N-1$,
to its
frequency representation
coefficients~$X[k]$,
$k = 0, 1, \ldots, N-1$,
according to~\cite[p.~543]{oppenheim1999discrete}:
\begin{align}
\label{equation-dft-definition}
    X[k]
    =
    \sum_{n=0}^{N-1}
    x[n] \cdot \omega_N^{n \cdot k}
    ,
    \qquad
    k = 0, 1, \ldots, N-1
    ,
\end{align}
where
$
\omega_N
=
\exp\left(
-
\frac{2\pi j}{N}
\right)
$
is the $N$th root of unity
and
$j \triangleq \sqrt{-1}$.
Under matrix formalism,
by letting
$\mathbf{x} = \begin{bmatrix} x[0] & x[1] & \cdots & x[N-1]\end{bmatrix}^\top$
and
$\mathbf{X} = \begin{bmatrix} X[0] & X[1] & \cdots & X[N-1]\end{bmatrix}^\top$,
we can write
\begin{align}
\mathbf{X}
=
\mathbf{F}_N \cdot \mathbf{x}
,
\end{align}
where $\mathbf{F}_N$ is
the DFT matrix of
order~$N$~\cite[p.~624]{oppenheim1999discrete},
whose~$(k,n)$th entry is given by
$\omega_{N}^{kn}$,
$k,n = 0, 1, \ldots, N-1$.
We adopt
the zero-based index
for matrix row and column count
(e.g., the first row index is $k=0$).
Matrix~$\mathbf{F}_N$
is hereafter referred to as
the standard DFT
as opposed to
the unitary DFT,
which is given by
$
\frac{1}{\sqrt{N}}
\cdot
\mathbf{F}_N
$~\cite{seber2008matrix}.

A DFT approximation---or approximate DFT (ADFT)---is a
matrix~$\widehat{\mathbf{F}}_N$ that
(i)~preserves relevant, context-dependent
mathematical properties of
the exact matrix~$\mathbf{F}_N$;
(ii)~exhibits performance
measurements close to
that of~$\mathbf{F}_N$
according to an appropriate
figure of merit;
and
(iii)~possesses low arithmetic complexity in
terms of the arithmetic operations count,
ensured by a numerical domain of
trivial multipliers,
such as
$
\big\{
0,
\pm 1,
\pm j,
\pm (1+j),
\pm (1-j)
\big\}
$~\cite{blahut2010}.

The task of
obtaining
an approximation~$\widehat{\mathbf{F}}_{N}$
that satisfies the above requirements
can be formulated as the
following optimization problem~\cite{portella2025multiplierless}:
\begin{align}
\label{eq-opt-general}
\widehat{\mathbf{F}}_N^\ast
=
\arg
\min_{\widehat{\mathbf{F}}_N}
\operatorname{error}
\left(
\mathbf{F}_N
,
\widehat{\mathbf{F}}_N
\right)
,
\end{align}
where
$\operatorname{error}(\cdot, \cdot)$
is an error measure
linked to the application context
(e.g., the Frobenius norm~\cite[p.~15]{seber2008matrix})
and
$
\widehat{\mathbf{F}}_N
$
is a candidate approximation for the DFT matrix.

\subsection{%
State-of-the-art
32-point Approximate DFT
Methods}

Among the state-of-the-art 32-point DFT approximations,
the following methods stand out:
(i)~the rounding-based approximation~\cite{cintra2011integer},
(ii)~the Cooley--Tukey-based approximation~\cite{coelho2024discrete},
and
(iii)~the Villagr\'an--Cintra
ADFT~\cite{suarezvillagran2015aproximacoes,cintra2024approximation}.
The latter method provides
superior performance
and has been mapped into
dedicated hardware
for application in beamforming systems~\cite{portella2022radix}.
For this reason,
we adopt the Villagr\'an--Cintra approximation
as the reference (\emph{baseline}) method
for subsequent comparisons with
the method
developed in this work.
Hereafter
the
Villagr\'an--Cintra design~\cite{cintra2024approximation}
is
denoted by~$\widehat{\mathbf{F}}^{(0)}_{32}$.
A detailed description
of~$\widehat{\mathbf{F}}^{(0)}_{32}$
is provided in~\cite{cintra2024approximation},
including its
definition,
error performance,
arithmetic complexity analysis,
and
fast algorithm.
Indeed,
the approximate
matrix~$\widehat{\mathbf{F}}^{(0)}_{32}$
admits
the following factorization:
\begin{align}
\label{eq:dora-cintra}
\widehat{\mathbf{F}}^{(0)}_{32}
=
\mathbf{W}_8
\cdot
\mathbf{W}_7
\cdot
\mathbf{W}_6
\cdot
\mathbf{W}_5
\cdot
\mathbf{W}_4
\cdot
\mathbf{W}_3
\cdot
\mathbf{W}_2
\cdot
\mathbf{W}_1
,
\end{align}
where
the
factors~$\mathbf{W}_i$,
$i = 1, 2, \ldots, 8$,
have explicit numerical expressions
available in~\cite{cintra2024approximation}.

\subsection{Goal}

In this work,
our objective is to derive
a competing 32-point DFT approximation
defined over
the set of trivial, complex-valued
multipliers
given by
$
\mathcal{P}[j]
\triangleq
\mathcal{P} + j \cdot \mathcal{P}
$,
where
$\mathcal{P} = \left\{0, \pm \frac{1}{2}, \pm 1 \right\}$.
To that end,
we aim at
addressing the optimization problem in~\eqref{eq-opt-general}.

\section{Methodology}

\subsection{Integer Optimization Problem}

In order to generate
candidate approximations
for the optimization problem
in~\eqref{eq-opt-general},
we adopt the following
characterization
of the approximate DFT (standard form)
suggested in~\cite{cintra2011integer}
and further developed in~\cite{portella2025multiplierless}.
Let~$\mathbf{T}$
be an integer matrix
with entries defined over
a set of trivial multipliers.
Thus,
a candidate approximation for
the DFT matrix is given by:
\begin{align}
\label{equation-polar}
\widehat{\mathbf{F}}_N
=
\sqrt{N}
\cdot
\operatorname{polar}
\left(\mathbf{T}
\right)
,
\end{align}
where
\begin{align}
\label{equation_polar_definition}
\operatorname{polar}
(\mathbf{T})
=
\sqrt{
    \left[
        \operatorname{diag}
            (
            \mathbf{T}
            \cdot
            \mathbf{T}^H
            )
    \right]^{-1}
    }
\cdot
\mathbf{T}
,
\end{align}
the operator~$\operatorname{diag}(\cdot)$
sets to zero the elements outside
the main diagonal of its matrix argument;
the superscript ${}^H$
denotes the Hermitian transposition,
and
the symbol~$\sqrt{\cdot}$
represents
the matrix square
root~\cite[p.~74]{seber2008matrix}.

The matrix
function~$\operatorname{polar}(\cdot)$
performs
a near-orthogonalization~\cite{cintra2011integer}
procedure
according to
the polar decomposition~\cite[p.~348]{seber2008matrix}
of its matrix argument
$\mathbf{T}$.
If
$
\mathbf{T}
\cdot
\mathbf{T}^\top
$
results in
a diagonal matrix,
then
$\operatorname{polar}(\mathbf{T})$
is a unitary matrix~\cite[p.~147]{seber2008matrix}.
Otherwise,
the approximation is not unitary,
but can be quasi-orthogonal
whenever
its deviation from orthogonality~\cite{flury1986algorithm}
is regarded as small.
Thus,
by means of the polar function,
the problem of finding
a competing DFT approximation
becomes
an integer optimization problem,
which consists of optimizing
the choice of
the low-complexity matrix~$\mathbf{T}$.

Nevertheless,
solving~\eqref{eq-opt-general}
by means of
exhaustive search is
computationally feasible
for small values of~$N$ only.
Indeed,
for $N=32$,
the search space consists of
$
\big[
\operatorname{card}(\mathcal{P}{[j]})
\big]^{32^2}
=
25^{32^2}
\approx
\num{3.09434604738257827548018e1431}
$
candidate matrices~$\mathbf{T}$,
where
$\operatorname{card}(\cdot)$
denotes
the cardinality of a set~\cite[p.~170]{rosen2019discrete}.

\subsection{Greedy Approach}

One way to address the large size of the search space
is the adoption of a
greedy algorithm approach~\cite[p.~414]{cormen2009introduction}.
First, we decompose the exact
matrix~$\mathbf{F}_{N}$ into its
real and imaginary parts;
next, each part is analyzed in a row-by-row manner,
so that
each row is sought to be independently  approximated.
In symbols,
we have the following formalism.
Let
$
\mathbf{c}_k
$
and
$
\mathbf{s}_k
$
denote row-vectors containing
the real and imaginary parts of the
$k$th row of $\mathbf{F}_N$,
respectively.
Therefore,
the components of
$
\mathbf{c}_k
$
and
$
\mathbf{s}_k
$
are,
respectively,
given by:
\begin{align}
c_{k,n} &
=
\Re\left\{ \omega_N^ {kn} \right\}
=
\cos\left(\frac{2\pi kn}{N}\right)
,
\qquad
n=0,1,\ldots,N-1
,
\\
s_{k,n} &
=
\Im\left\{ \omega_N^ {kn} \right\}
=
- \sin\left(\frac{2\pi kn}{N}\right)
,
\qquad
n=0,1,\ldots,N-1
.
\end{align}
where
$\Re\{ \cdot \}$
and
$\Im\{ \cdot \}$
return
the real and imaginary parts of
their arguments,
respectively.

We seek to replace
$\mathbf{c}_k$ and $\mathbf{s}_k$
with row-vector approximations
$
\hat{\mathbf{c}}_k
$
and
$
\hat{\mathbf{s}}_k
$.
The approximate rows
are dependent
on a low-complexity vector
$\mathbf{t}$
and
can
be obtained by solving
the following minimization problems:
\begin{align}
\Re\left\{ \mathbf{t}^\ast_k \right\}
=
\arg
\min_{\mathbf{t} \in \mathcal{P}^N}
&
\|
\mathbf{c}_k
-
\widehat{\mathbf{c}}_k
\|_{F}
,
\qquad
k = 0, 1, \ldots, N-1
,
\\
\Im\left\{ \mathbf{t}^\ast_k \right\}
=
\arg
\min_{\mathbf{t} \in \mathcal{P}^N}
&
\|
\mathbf{s}_k
-
\widehat{\mathbf{s}}_k
\|_{F}
,
\qquad
k = 0, 1, \ldots, N-1
,
\end{align}
where
$\| \cdot \|_F$
denotes the Frobenius
norm~\cite[p.~15]{seber2008matrix}.

\subsection{Search Space Reduction}

A way to substantially further reduce the size
of the search space
is to ensure
that the sought approximation matrix
possesses
some of the
symmetry patterns exhibited by the DFT itself.
To this end,
we impose
two conditions on the components of the
candidate vectors:
\begin{enumerate}[(i)]
\item
elements of the same absolute value
in the exact matrix must
effect
approximate elements of the same absolute value;

\item
the approximate matrix must preserve the sign pattern
of the exact matrix elements.

\end{enumerate}
Symbolically,
these two criteria
correspond
to:
(i)~if
$|c_{k,n}| = |c_{k,m}|$, $n\neq m$,
then
$
|\widehat{c}_{k,n}|
=
|\widehat{c}_{k,m}|
$;
and
(ii)~$\operatorname{sign}(c_{k,n}) =
\operatorname{sign}(\widehat{c}_{k,n})$.
Analogous considerations
apply to the elements~$s_{k,n}$,
$n=0,1,\ldots,N-1$.
A row-vector satisfying the above
conditions is hereafter referred to as \emph{DFT-consistent}.
Therefore,
for a given DFT row~$k$,
we have
the space of low-complexity,
DFT-consistent
as the set of
$N$-point vectors in $\mathcal{P}[j]^N$
described
as follows:
\begin{align}
\mathcal{S}_k
=
\big\{
\mathbf{t}
\colon
\mathbf{t}
\in
\mathcal{P}[j]^N
,
\text{$\mathbf{t}$ is DFT-consistent with the $k$th row of $\mathbf{F}_N$}
\big\}
,
\qquad
k = 0, 1, \ldots, N-1
.
\end{align}

For example,
for $k=1$,
the 8-point DFT matrix
($N=8$)
presents the following row:
\begin{align}
\begin{bmatrix}
1 &
\frac{\sqrt{2}}{2} - \frac{\sqrt{2}}{2}j &
-j &
-\frac{\sqrt{2}}{2} - \frac{\sqrt{2}}{2}j &
-1 &
-\frac{\sqrt{2}}{2} + \frac{\sqrt{2}}{2}j &
j &
\frac{\sqrt{2}}{2} + \frac{\sqrt{2}}{2}j
\end{bmatrix}^\top
\end{align}
Therefore,
we have the following 2-parameter mapping:
\begin{align}
\begin{bmatrix}
    a_0 & a_1
\end{bmatrix}^\top
\mapsto
\begin{bmatrix}
a_0 &
a_1 - j\cdot a_1 &
-j \cdot a_0 &
-a_1 - j\cdot a_1 &
-a_0 &
-a_1 + j\cdot a_1 &
j \cdot a_0 &
a_1 + j\cdot a_1
\end{bmatrix}^\top
.
\end{align}
Taking the real and imaginary parts,
the above mapping
can be split
into two functions
as follows:
\begin{align}
\operatorname{C}_{1,8}(a_0, a_1)
&=
\begin{bmatrix}
|a_0| &
|a_1|  &
0 &
-|a_1| &
-|a_0| &
-|a_1| &
0 &
|a_1|
\end{bmatrix}^\top
,
\\
\operatorname{S}_{1,8}(a_0, a_1)
&=
\begin{bmatrix}
0 &
-|a_1| &
-|a_0| &
-|a_1| &
0 &
|a_1| &
|a_0| &
|a_1|
\end{bmatrix}^\top
.
\end{align}
Thus,
by running the variables
$a_0$ and $a_1$ through
the set
$\mathcal{P}$,
we obtain all possible DFT-consistent,
low-complexity vectors for this particular case ($k=1$, $N=8$).
Such a structure defines a suitable
search space of
candidate vectors.

In general,
vector functions
$\operatorname{C}_{k,N}(\cdot)$
and
$\operatorname{S}_{k,N}(\cdot)$
can be directly obtained
by inspecting
the elements in the $k$th row
of
the exact $N$-point DFT matrix.
Thus,
it is a straightforward
procedure
to generate the following
sets:
\begin{align}
\Re\left\{\mathcal{S}_k\right\}
&=
\big\{
\mathbf{t}
\colon
\mathbf{t} = \operatorname{C}_{k,N}(\mathbf{a})
,
\mathbf{a} \in \mathcal{P}^{u_k}
\big\}
,
\\
\Im\left\{\mathcal{S}_k\right\}
&
=
\big\{
\mathbf{t}
\colon
\mathbf{t} = \operatorname{S}_{k,N}(\mathbf{a})
,
\mathbf{a} \in \mathcal{P}^{v_k}
\big\}
,
\end{align}
where
$u_k$ and $v_k$
are
the number of parameters.

By examining the uniqueness
in absolute value
of the elements
in the DFT matrix,
one can
determine $u_k$ and $v_k$.
From the algebraic
properties of
the $N$th root of the unity~\cite{herstein1975},
the number of unique values
of
$\omega_N^k$,
$k=1,2,\ldots,N-1$,
is given by
\begin{align}
l_k
=
\frac{N}{4}
\cdot
\frac{1}{\gcd(k,8)}
=
\frac{1}{k} \cdot \operatorname{lcm}\left(\frac{N}{4},k \right)
,
\end{align}
where
$\gcd(\cdot,\cdot)$
is the greatest common divisor~\cite[p.~21]{burton2011elementary}
and
$\operatorname{lcm}(\cdot,\cdot)$ is
the least common multiple~\cite[p.~29]{burton2011elementary}.
Let
the operator
$\operatorname{unique}(\mathbf{x})$
return
the number of nonnull unique elements
of an input vector~$\mathbf{x}$.
Thus,
we define the quantities:
\begin{align}
u_k
&=
\operatorname{unique}( |\mathbf{c}_k| )
\leq
l_k
,
\\
v_k
&=
\operatorname{unique}( |\mathbf{s}_k| )
\leq
l_k
,
\end{align}
$k = 0, 1, \ldots, N-1$.

To further reduce the computational
cost of solving this problem,
during the search phase,
we adopted
the simplification:
$
\widehat{\mathbf{F}}_N
=
\mathbf{T}
$.
Then,
we have the following
constrained optimization
problems:
\begin{align}
\label{eq-min-ck}
\Re\left\{
\mathbf{t}^\ast_k
\right\}
=
\arg
\min_{\mathbf{t} \in \Re\{\mathcal{S}_k\}}
&
\|
\mathbf{c}_k
-
\mathbf{t}
\|_{F},
\qquad
k = 0, 1, \ldots, N-1
,
\\
\label{eq-min-sk}
\Im\left\{
\mathbf{t}^\ast_k
\right\}
=
\arg
\min_{\mathbf{t} \in \Im\{\mathcal{S}_k\}}
&
\|
\mathbf{s}_k
-
\mathbf{t}
\|_{F}
,
\qquad
k = 0, 1, \ldots, N-1
.
\end{align}
Finally,
the $k$th row
of the sought low-complexity matrix~$\mathbf{T}^\ast$
consists of
$
\mathbf{t}^\ast_k
=
\Re\left\{
\mathbf{t}^\ast_k
\right\}
+
j \cdot
\Im\left\{
\mathbf{t}^\ast_k
\right\}
$.

\section{Results}

\subsection{Low-complexity Matrix}

In this section,
we apply the discussed methodology
for the case $N=32$
and
$\mathcal{P} = \left\{0, \frac{1}{2}, 1 \right\}$.
Consequently,
we obtain:
\begin{align}
u_k = v_k
&=
\begin{cases}
1, & \text{if $8 \mid k$}
,
\\
\frac{1}{k} \cdot \operatorname{lcm}(8,k)
, & \text{otherwise.}
\end{cases}
\end{align}
The separate case when
$8 \mid k$
is due to the fact that
the 0th, 8th, 16th, and 24th
rows of the 32-point DFT matrix
consist of trivial multipliers
$\pm1$, $\pm j$.
Thus,
for such rows,
there is no approximation problem to address.
The sizes of the search space of the minimization
problems in~\eqref{eq-min-ck} and \eqref{eq-min-sk}
are
given by:
$
\sum_{k=1}^{32}
[\operatorname{card}(\mathcal{P})]^{u_k}
 =
\sum_{k=1}^{32}
[\operatorname{card}(\mathcal{P})]^{v_k}
=
\num[round-precision   = 0]{105664}
$.
Therefore,
a total of
\num{211.3e3}
candidate vectors,
which is a comparatively small number,
rendering
the problem tractable.
Thus,
the minimization problems
in~\eqref{eq-min-ck} and~\eqref{eq-min-sk},
for $N=32$,
could be given a solution
by means of exhaustive search.
The
obtained solution is
referred to as~$\mathbf{T}^\ast_{32}$.

The proposed low-complexity matrix
$\mathbf{T}^\ast_{32}$
can be compactly represented as
shown below:
\begin{align}
\mathbf{T}^\ast_{32}
=
\frac{1}{2}\cdot
\begin{bmatrix}
\mathbf{A} & \mathbf{D} \cdot \mathbf{A} \\
\mathbf{A}\cdot\mathbf{D} & \mathbf{D} \cdot \mathbf{A} \cdot \mathbf{D}
\end{bmatrix},
\label{eq:block_F32}
\end{align}
where
$
\mathbf{D}
=
\operatorname{diag}\left(1,-1,1,-1,\ldots,1,-1\right)
$
is $16\times 16$
and
\begin{align}
	\mathbf{A}
	=
	\begin{bsmallmatrix*}[r]
		2 & 2 & 2 & 2 & 2 & 2 & 2 & 2 & 2 & 2 & 2 & 2 & 2 & 2 & 2 & 2 \\
		2 & 2 & 2-j & 2-j & 1-j & 1-2j & 1-2j & -2j & -2j & -2j & -1-2j & -1-2j & -1-j & -2-j & -2-j & -2 \\
		2 & 2-j & 1-j & 1-2j & -2j & -1-2j & -1-j & -2-j & -2 & -2+j & -1+j & -1+2j & 2j & 1+2j & 1+j & 2+j \\
		2 & 2-j & 1-2j & -2j & -1-j & -2 & -2+j & -1+2j & 2j & 1+2j & 2+j & 2 & 1-j & -2j & -1-2j & -2-j \\
		2 & 1-j & -2j & -1-j & -2 & -1+j & 2j & 1+j & 2 & 1-j & -2j & -1-j & -2 & -1+j & 2j & 1+j \\
		2 & 1-2j & -1-2j & -2 & -1+j & 2j & 2+j & 2-j & -2j & -2-j & -2+j & 2j & 1+j & 2 & 1-2j & -1-2j \\
		2 & 1-2j & -1-j & -2+j & 2j & 2+j & 1-j & -1-2j & -2 & -1+2j & 1+j & 2-j & -2j & -2-j & -1+j & 1+2j \\
		2 & -2j & -2-j & -1+2j & 1+j & 2-j & -1-2j & -2 & 2j & 2 & 1-2j & -2-j & -1+j & 1+2j & 2-j & -2j \\
		2 & -2j & -2 & 2j & 2 & -2j & -2 & 2j & 2 & -2j & -2 & 2j & 2 & -2j & -2 & 2j \\
		2 & -2j & -2+j & 1+2j & 1-j & -2-j & -1+2j & 2 & -2j & -2 & 1+2j & 2-j & -1-j & -1+2j & 2+j & -2j \\
		2 & -1-2j & -1+j & 2+j & -2j & -2+j & 1+j & 1-2j & -2 & 1+2j & 1-j & -2-j & 2j & 2-j & -1-j & -1+2j \\
		2 & -1-2j & -1+2j & 2 & -1-j & 2j & 2-j & -2-j & 2j & 2-j & -2-j & 2j & 1-j & -2 & 1+2j & 1-2j \\
		2 & -1-j & 2j & 1-j & -2 & 1+j & -2j & -1+j & 2 & -1-j & 2j & 1-j & -2 & 1+j & -2j & -1+j \\
		2 & -2-j & 1+2j & -2j & -1+j & 2 & -2-j & 1+2j & -2j & -1+2j & 2-j & -2 & 1+j & -2j & -1+2j & 2-j \\
		2 & -2-j & 1+j & -1-2j & 2j & 1-2j & -1+j & 2-j & -2 & 2+j & -1-j & 1+2j & -2j & -1+2j & 1-j & -2+j \\
		2 & -2 & 2+j & -2-j & 1+j & -1-2j & 1+2j & -2j & 2j & -2j & -1+2j & 1-2j & -1+j & 2-j & -2+j & 2
	\end{bsmallmatrix*}
	.
\end{align}

\subsection{32-point DFT Approximations}

Based on $\mathbf{T}^\ast_{32}$,
we define three interrelated
approximations
for the 32-point DFT.
The approximate DFT matrices
are described below:

\begin{enumerate}[(i)]
\item
The raw approximate DFT
is given directly
by the low-complexity matrix:
\begin{align}
\hat{\mathbf{F}}_{32}^{(1)}
\triangleq
\mathbf{T}^\ast_{32}
.
\end{align}
This form might be useful
when
unscaled, unnormalized
spectral estimation is sufficient.
For example,
in the context
of hypothesis testing
for frequency-content detection,
the
threshold levels
can be adjusted
to compensate
the unnormalized
spectral estimates~\cite{kay1998fundamentals};

\item
The normalized approximate DFT
results from the near-orthogonalization
process~\cite{cintra2011integer,portella2025multiplierless}
implied by
the polar
decomposition~\cite[p.~348]{seber2008matrix}:
\begin{align}
\hat{\mathbf{F}}_{32}^{(2)}
\triangleq
\operatorname{polar}\big( \mathbf{T}^\ast_{32} \big)
;
\end{align}
This form finds application
when the unitary DFT
is required~\cite[p.~184]{seber2008matrix}.

\item
The standard approximate DFT
is obtained
from the normalized approximation
according to
the multiplication
by the scaling factor~$\sqrt{N}$.
For the 32-point case,
we have:
\begin{align}
\hat{\mathbf{F}}_{32}^{(3)}
\triangleq
\sqrt{32}
\cdot
\hat{\mathbf{F}}_{32}^{(2)}
.
\end{align}
This form furnishes an approximate DFT that works
in close agreement to
the standard DFT~\cite[p.~543]{oppenheim1999discrete}
(cf.~\eqref{equation-dft-definition}).

\end{enumerate}

\subsection{Factorization: Fast Algorithm}

The obtained low-complexity matrix~$\mathbf{T}^\ast_{32}$
admits the following factorization into sparse matrices:
\begin{align}
\label{equation-factorization}
\mathbf{T}^\ast_{32}
=
\frac{1}{2}
\cdot
\mathbf{W}_{13}
\cdot
\mathbf{D}_3
\cdot
\mathbf{W}_{12}
\cdot
\mathbf{W}_{11}
\cdot
\mathbf{W}_{10}
\cdot
\mathbf{D}_2
\cdot
\mathbf{W}_{9}
\cdot
\mathbf{D}_1
\cdot
\mathbf{W}_3
\cdot
\mathbf{W}_2
\cdot
\mathbf{W}_1
,
\end{align}
where
the matrix factors
$\mathbf{W}_1$,
$\mathbf{W}_2$,
and
$\mathbf{W}_3$
are inherited from
the factorization
of~$\widehat{\mathbf{F}}_{32}^{(0)}$ in~\eqref{eq:dora-cintra}.
The matrix factors
$\mathbf{W}_i$,
$i = 9,10,11,12,13$,
are
described in compact notation
as
\begin{align}
\mathbf{W}_{9} &= \mathbf{P}_{9}^\top \cdot
  \Big[
    (\mathbf{I}_{2} \otimes \mathbf{B}_2) \oplus 1
    \oplus \mathbf{C}_{2}
      \oplus \mathbf{C}_{1}
      \oplus \mathbf{C}_{2}
      \oplus \mathbf{C}_{1}
      \oplus
      \mathbf{B}_2 \oplus{}
    \mathbf{I}_{2} \oplus
    \mathbf{C}_{3} \oplus
    \mathbf{C}_{2}
      \oplus \mathbf{C}_{1}
      \oplus \mathbf{C}_{3}
      \oplus \mathbf{I}_{3}
  \Big]
  \cdot \mathbf{P}_{9}
  ,
  \\
\mathbf{W}_{10}&= \mathbf{P}_{10}^\top \cdot
  \Big[
    \mathbf{B}_2 \oplus 1 \oplus
    \mathbf{C}_{1} \oplus
    (\mathbf{I}_{4} \otimes \mathbf{B}_2)
      \oplus \mathbf{C}_{1}
      \oplus 1 \oplus{}
    \mathbf{B}_2 \oplus 1 \oplus
    \mathbf{B}_2 \oplus \mathbf{I}_{2}
    \oplus \mathbf{B}_2
      \oplus \mathbf{I}_{2} \oplus
      \mathbf{C}_{2} \oplus
      \mathbf{I}_{3}
  \Big]
  \cdot \mathbf{P}_{10}
  ,
  \\
\mathbf{W}_{11} &= \mathbf{P}_{11}^\top \cdot
  \Big[\mathbf{B}_2 \oplus \mathbf{I}_{14}
  \oplus (\mathbf{I}_{5} \otimes \mathbf{B}_2)
    \oplus \mathbf{I}_{2} \oplus
    \mathbf{B}_2 \oplus \mathbf{I}_{2}\Big]
  \cdot \mathbf{P}_{11}
  ,
  \\
\mathbf{W}_{12} &= \mathbf{P}_{12}^\top \cdot
  \Big[\mathbf{I}_{16} \oplus \mathbf{B}_2
  \oplus \mathbf{I}_{8}
    \oplus (\mathbf{I}_{3} \otimes
    \mathbf{B}_2)\Big]
  \cdot \mathbf{P}_{12}
  ,
  \\
\mathbf{W}_{13} &=
\begin{bmatrix}
2 \oplus \mathbf{I}_{15} & 0 \oplus \mathbf{I}_{15}
\\
0 \oplus \mathbf{J}_{15} & 2 \oplus (-\mathbf{J}_{15})
\end{bmatrix}
\cdot
\left(
\mathbf{I}_{17}
\oplus
(j\cdot \mathbf{I}_{15})
\right)
\cdot
\mathbf{D}_4
\cdot \mathbf{P}_{13},
\end{align}
where
$\otimes$
and
$\oplus$
denote
the Kronecker product and
the direct sum~\cite[p.~234]{seber2008matrix},
respectively,
$\mathbf{I}_m$ is
the $m \times m$ identity matrix,
and
$\mathbf{J}_m$ is
the $m \times m$
exchange matrix~\cite[p.~159]{seber2008matrix},
the diagonal matrices
$\mathbf{D}_i$,
$i = 1,2,3,4,$
are given by
\begin{align}
    \mathbf{D}_1
    &=
    \mathbf{I}_{8} \oplus 2 \oplus \mathbf{I}_{3}
    \oplus 2 \oplus \mathbf{I}_{11} \oplus 2 \oplus \mathbf{I}_{7}
    ,
    \\
    \mathbf{D}_2
    &=
    \mathbf{I}_{4} \oplus 2 \oplus \mathbf{I}_{9}
    \oplus 2 \oplus 1 \oplus 2 \oplus \mathbf{I}_{15}
    ,
    \\
    \mathbf{D}_3
    &=
    \mathbf{I}_{2} \oplus 2 \oplus \mathbf{I}_{12}
    \oplus 2 \oplus \mathbf{I}_{16}
    ,
    \\
    \mathbf{D}_4
&=
\mathbf{I}_{5} \oplus -1
\oplus \mathbf{I}_{4}
    \oplus -1 \oplus
    \mathbf{I}_{2} \oplus
    -\mathbf{I}_{2}
    \oplus
    \mathbf{I}_{3}
    \oplus
    -1 \oplus \mathbf{I}_{4}
\oplus -\mathbf{I}_{2}
\oplus \mathbf{I}_{3}
\oplus -1 \oplus 1
\oplus -1 \oplus 1
    ,
\end{align}
the matrix factors~$\mathbf{P}_{i}$,
$i = 9,10,11,12,13$,
are permutation matrices
compactly
expressed in
cycle notation~\cite[p.~76]{herstein1975}
in Table~\ref{tab:permutations},
and
the auxiliary submatrices are:
\begin{align*}
\mathbf{B}_2 =
\begin{bmatrix*}[r]
    1 & 1 \\
    1 & -1
\end{bmatrix*}
,
\quad
\mathbf{C}_{1} =
\begin{bmatrix*}[r]
    1 & 1
    \\
    -1 & 1
\end{bmatrix*}
,
\quad
\mathbf{C}_{2} =
\begin{bmatrix*}[r]
    1 & 2 \\
    -2 & 1
\end{bmatrix*}
,
\quad
\text{and}
\quad
\mathbf{C}_{3} =
\begin{bmatrix*}[r]
    0 & -1 & -2 & -2 \\
    -1 & -2 & 0 & 2 \\
    -2 & 0 & 2 & -1 \\
    -2 & 2 & -1 & 0
\end{bmatrix*}
.
\end{align*}

\begin{table}
\centering
\caption{Permutation Matrices in Cycle Notation}
\label{tab:permutations}
\begin{tabular}{
    c
    l
    }
    \toprule
    Permutation & Cycle notation \\
    \midrule
    $\mathbf{P}_{9}$ &
    $(2\;5\;3)$  $(7\;8)$ $(11\;12)$ $(15\;16)$ $(21\;24)$
    $(19\;20\;22)$ $(27\;28\;30)$ $(29\;32\;31)$
    \\
    \midrule
    $\mathbf{P}_{10}$ &
    $(2\;3)$ $(18\;29\;31\;30\;28\;27\;26\;24\;23\;22\;20\;19)$
    \\
    \midrule
    $\mathbf{P}_{11}$ &
    $(18\;27\;26\;25\;22\;23\;21\;20\;19)$ $(30\;31)$
    \\
    \midrule
    $\mathbf{P}_{12}$ &
    $(18\;26\;25\;24\;23\;22\;21\;20\;19)$ $(28\;32\;31\;30\;29)$
    \\
    \midrule
    $\mathbf{P}_{13}$ &
    $(2\;26\;21\;14\;29\;15\;7\;8\;27\;13\;5\;4\;28\;25\;16\;17)$
    $(3\;6\;31\;11\;9)$  $(10\;32\;18\;19)$  $(12\;30\;23)$ \\

    \bottomrule
\end{tabular}
\end{table}

To facilitate reproducibility,
the symbolically described matrix
terms
in~\eqref{equation-factorization}
are given
explicit numerical forms
in the Appendix.

\section{Assessment and Comparison}

In this section,
the proposed approximations
$\hat{\mathbf{F}}_{32}^{(1)}$
and
$\hat{\mathbf{F}}_{32}^{(3)}$
are compared with
the standard DFT
matrix~$\mathbf{F}_{32}$
and
the baseline method
$\hat{\mathbf{F}}_{32}^{(0)}$;
whereas
the proposed approximation
$\hat{\mathbf{F}}_{32}^{(2)}$
is compared
with
the unitary DFT
matrix~$
\frac{1}{\sqrt{32}}
\cdot
\mathbf{F}_{32}
$.

\subsection{Error Analysis}

Table~\ref{table-energy-error-matrix}
shows the
results from the
performance assessment
according
to traditional
figures of merit in the field:
(i)~the total energy error
denoted
by
$
\operatorname{\varepsilon}
$~\cite{cintra2011dct};
(ii)~the mean absolute percentage error (MAPE)~\cite{Hyndman2018};
and
(iii)~the
deviation from orthogonality
represented
by~$
\operatorname{\delta}
$~\cite{flury1986algorithm}.
The measurements are taken relative
to the exact DFT.

\begin{table}
\centering
\caption{Performance Evaluation for
$\widehat{\mathbf{F}}_{32}^{(0)}$,
$\widehat{\mathbf{F}}_{32}^{(1)}$,
$\widehat{\mathbf{F}}_{32}^{(2)}$,
and
$\widehat{\mathbf{F}}_{32}^{(3)}$}
\label{table-energy-error-matrix}
\begin{tabular}{
    l
    S[round-precision = 2]
    S[round-precision = 2]
    S[round-precision = 1, exponent-mode = scientific]
    }
\toprule
Method &
$\operatorname{\varepsilon}$ &
$\operatorname{MAPE}$
& $\operatorname{\delta}$
\\
\midrule
$\widehat{\mathbf{F}}_{32}^{(0)}$~\cite{cintra2024approximation}
& 332.230 & 25.913 & 0.034
\\ [8pt]
$\widehat{\mathbf{F}}_{32}^{(1)}$ (proposed; raw) &
95.591 & 13.657 & 0.041
\\ [8pt]
$\widehat{\mathbf{F}}_{32}^{(2)}$ (proposed; normalized) &
2.714 & 13.633 & 0.041
\\ [8pt]
$\widehat{\mathbf{F}}_{32}^{(3)}$ (proposed; standard) &
86.857 & 13.632 & 0.041
\\
\bottomrule
\end{tabular}
\end{table}

\subsection{Frequency Analysis}

A transformation matrix
can be interpreted
as a filter bank
where each matrix row defines
a finite impulse response (FIR) filter~\cite{oppenheim1999discrete}.
Thus,
an approximate transform can be assessed
by comparing the frequency responses
from the implied
exact and approximate filters~\cite{cintra2011integer}.
We adopt the following notation.
Let
$
f_k[n]
$
and
$
\hat{f}_k[n]
$,
$n= 0,1,\ldots,N-1$,
denote
discrete-time signals
defined by
the $k$th row of
the exact and the approximate DFT
matrices,
respectively.
The frequency responses
are obtained from the discrete-time Fourier transform
of
$f_k[n]$ and $\hat{f}_k[n]$
,
which are
denoted
by
$F(\omega; k)$
and
$\hat{F}(\omega; k)$,
where
$0 \leq \omega < 2\pi$
is
the radian frequency~\cite[p.~40]{oppenheim1999discrete}.
In general,
the total energy error between two filters
can be obtained by
the energy contents
of the difference between their
frequency responses~\cite[p.~51]{oppenheim1999discrete}.
For instance,
the energy error
between
$F(\omega; k)$
and
$\hat{F}(\omega; k)$
is:
\begin{align}
\epsilon_k
\big(
F(\omega; k), \hat{F}(\omega; k)
\big)
=
\frac{1}{2\pi}
\int_{-\pi}^{\pi}
|
F(\omega; k) - \hat{F}(\omega; k)
|^2
\operatorname{d}\omega
=
\sum_{n=0}^{N-1}
\left|
f_k[n] - \hat{f}_k[n]
\right|^2
,
\label{eq:energy_error}
\end{align}
being the last equality
due to
Parseval's
theorem~\cite[p.~60]{oppenheim1999discrete}.

Next,
we adopt the following notation:
$\hat{F}^{(0)}(\omega;k)$,
$\hat{F}^{(1)}(\omega;k)$,
$\hat{F}^{(2)}(\omega;k)$,
and
$\hat{F}^{(3)}(\omega;k)$
correspond
to the frequency responses
associated with the $k$th row
of
the
baseline method
$\hat{\mathbf{F}}_{32}^{(0)}$
and
the proposed methods:
$\widehat{\mathbf{F}}_{32}^{(1)}$,
$\widehat{\mathbf{F}}_{32}^{(2)}$,
and
$\widehat{\mathbf{F}}_{32}^{(3)}$,
respectively.
For the 32-point case,
we notice that
rows $k\in\{0,8,16,24\}$ of the exact DFT
consist of trivial multipliers;
therefore
they yield null energy error.
The numerical evaluation
of the energy error
reveals the figures shown in Table~\ref{table-energy-error-by-row}.
Although
the measurements in
Table~\ref{table-energy-error-by-row}
are given in decimal format,
they can be given
exact expressions.
For instance,
routine algebraic manipulation
yields:
\begin{align}
\operatorname{\epsilon}
\left(
F(\omega ; 0), \hat{F}^{(1)}(\omega;0)
\right)
&=
0
\qquad
\text{(exact)},
\\
\operatorname{\epsilon}
\left(
F(\omega ; 1), \hat{F}^{(1)}(\omega;1)
\right)
&=
8 \times
\left(
\frac{29}{4}
-\frac{\sqrt{2}}{2}
-2(\alpha_1+\alpha_2+\alpha_3)-(\beta_2+\beta_3)
\right)
,
\\
\operatorname{\epsilon}
\left(
F(\omega ; 2), \hat{F}^{(1)}(\omega;2)
\right)
&=
16 \times
\left(
3 - \frac{\sqrt{2}}{2} - 2\alpha_2 -\beta_2
\right)
,
\\
\operatorname{\epsilon}
\left(
F(\omega ; 4), \hat{F}^{(1)}(\omega;4)
\right)
&=
32 \times \left(\frac{3}{4} -\frac{\sqrt{2}}{2}\right)
,
\end{align}
where
$
\alpha_1 =
\frac{1}{2}\sqrt{2+\sqrt{2+\sqrt{2}}}
$,
$
\beta_1 =
\frac{1}{2}\sqrt{2-\sqrt{2+\sqrt{2}}}
$,
$\alpha_2 = \frac{1}{2}\sqrt{2+\sqrt{2}}$,
$\beta_2 = \frac{1}{2}\sqrt{2-\sqrt{2}}$,
$\alpha_3 = \alpha_1 \alpha_2 - \beta_1 \beta_2$,
and
$\beta_3 = \alpha_1 \beta_2 + \beta_1 \alpha_2$.
Similar algebraic derivations
can be done for the other cases.

\begin{table}[]
    \centering
\caption{Energy Error between Frequency Responses}
    \label{table-energy-error-by-row}

\sisetup{
        round-precision   = 3,
        exponent-mode = threshold,
        exponent-thresholds = -2:2,
}

\begin{tabular}{
                c|
                S
                @{}S
                @{}S
                @{}S
                }
\toprule
Energy Error    &
{$k \equiv 0 \pmod{8}$}   &
{$k \equiv 1 \pmod{2}$}   &
{$k \equiv 2 \pmod{4}$}   &
{$k \equiv 4 \pmod{8}$}
\\
\midrule
$\operatorname{\epsilon}\left( F(\omega ; k), \hat{F}^{(0)}(\omega;k) \right)$ &
0   & 4.01901696 & 3.80843796 &  2.74516600
\\ [8pt]
$\operatorname{\epsilon}\left( F(\omega ; k), \hat{F}^{(1)}(\omega;k) \right)$ &
 0 & 1.05896562 & 0.99921154 & 1.37258300
\\ [8pt]
$\operatorname{\epsilon}\left( \frac{F(\omega ; k)}{\sqrt{32}}, \hat{F}^{(2)}(\omega;k) \right)$
 & 0 & 0.0311857 & 0.0312253 & 0.0288028
\\ [8pt]
$\operatorname{\epsilon}\left( F(\omega ; k), \hat{F}^{(3)}(\omega;k) \right)$ &
 0 & 0.997942400000000  & 0.999209600000000  & 0.921689600000000
\\
\bottomrule
\end{tabular}

\end{table}

According to
the results shown in
Table~\ref{table-energy-error-by-row},
we separated the filters
with lowest nonnull (best cases)
and
highest (worst cases)
energy errors.
For instance,
considering
the baseline
approximation~$\widehat{\mathbf{F}}_{32}^{(0)}$,
the smallest nonzero error is
\num[round-precision=3]{2.74516600},
obtained for
$k \in \{ 4, 12, 20, 28 \}$ ($k\equiv4\pmod8)$.
The row $k=4$ was chosen as
the representative of the best case.
On the other hand,
the largest error is
\num[round-precision=3]{4.01901696},
obtained
for any odd row.
The row $k=1$ was adopted as
the representative of the worst case.
Any other odd row, such as $k=3$, would produce the same error.

\begin{figure}
    \centering
    \subfigure
    [$|\hat{F}^{(1)}(\omega; k)|$]
    {\includegraphics[scale=0.975]{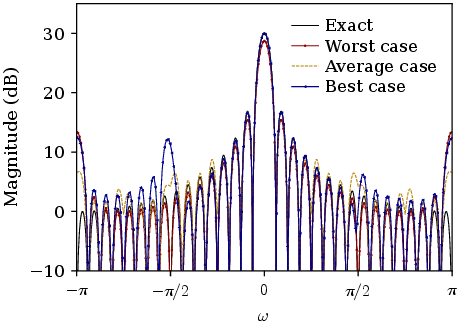}
    \label{fig:resp-exact-raw}}
    \subfigure
        [$\angle \hat{F}^{(1)}(\omega; k)$]
        {\includegraphics[scale=0.975]{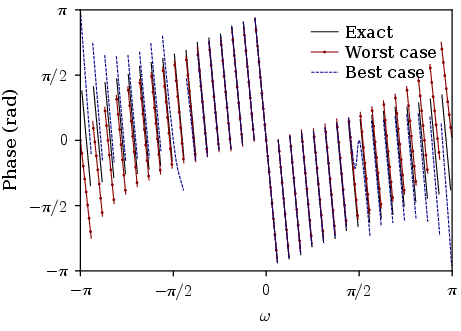}
            \label{fig:phase-exact-raw}
        }

    \subfigure
    [$|\hat{F}^{(2)}(\omega; k)|$]
    {\includegraphics[scale=0.975]{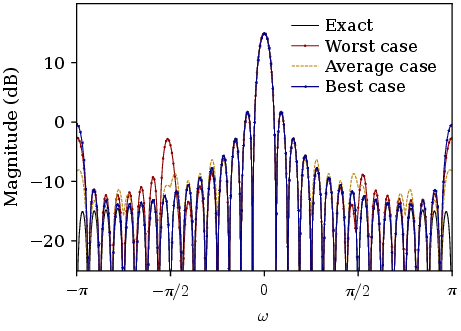}
    \label{fig:resp-exact-unitary}}
        \subfigure
        [$\angle \hat{F}^{(2)}(\omega; k)$]
        {%
            \includegraphics[scale=0.975]{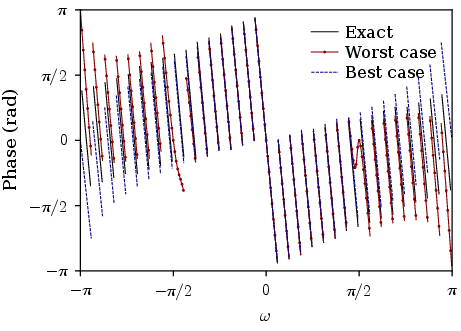}
            \label{fig:phase-exact-unitary}
        }

    \subfigure
    [$|\hat{F}^{(3)}(\omega; k)|$]
    {\includegraphics[scale=0.975]{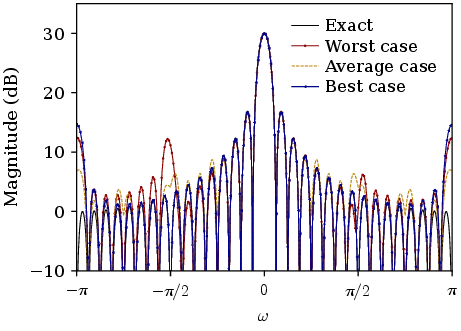}
    \label{fig:resp-exact-proposed}
    }
        \subfigure
        [$\angle \hat{F}^{(3)}(\omega; k)$]
        {%
            \includegraphics[scale=0.975]{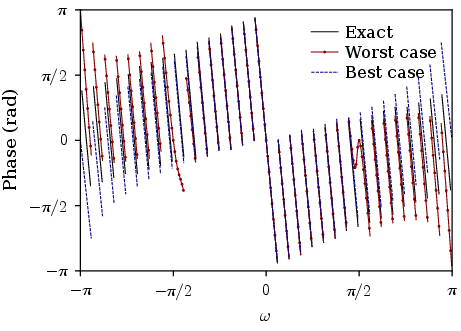}
            \label{fig:phase-exact-proposed}
        }

    \caption{%
    Frequency response
    magnitude (left column) and
    phase (right column)
    plots:
    (a)-(b)~Raw approximation:
    Best case $\hat{F}^{(1)}(\omega; 2)$,
    worst case $\hat{F}^{(1)}(\omega; 4)$;
    (c)-(d)~Normalized approximation:
    Best case $\hat{F}^{(2)}(\omega; 4)$,
    worst case $\hat{F}^{(2)}(\omega; 2)$;
    (e)-(f)~Standard approximation:
    Best case $\hat{F}^{(3)}(\omega; 4)$,
    worst case $\hat{F}^{(3)}(\omega; 2)$;
    }
    \label{fig:frequency-response-comparison}
\end{figure}

Figure~\ref{fig:frequency-response-comparison} presents
the frequency response magnitude and phase
plots
corresponding
to the three proposed approximations.
Thus,
we have:
(i)~for the raw approximation,
$\hat{F}^{(1)}(\omega; 2)$ (best case)
and
$\hat{F}^{(1)}(\omega; 4)$ (worst case)
(Figure~\ref{fig:frequency-response-comparison}(a)-(b));
(ii)~for the normalized approximation,
$\hat{F}^{(2)}(\omega; 4)$ (best case)
and
$\hat{F}^{(2)}(\omega; 2)$ (worst case)
(Figure~\ref{fig:frequency-response-comparison}(c)-(d));
and
(iii)~for the standard approximation,
$\hat{F}^{(3)}(\omega; 4)$ (best case)
and
$\hat{F}^{(3)}(\omega; 2)$ (worst case)
(Figure~\ref{fig:frequency-response-comparison}(e)-(f)).
The raw and standard forms are compared with the exact DFT
(cf.~\eqref{equation-dft-definition});
whereas
the normalized form is compared with
the exact unitary
DFT~\cite[p.~147]{seber2008matrix}.
For better visualization and qualitative comparison,
the plots were
superimposed and centralized.
We also include the average case
computed by taking the mean
of
the centralized frequency responses
across all filters.

\begin{figure}
    \centering

    \subfigure[Best case: $|\hat{F}^{(0)}(\omega; 4)|$ and $|\hat{F}^{(1)}(\omega; 2)|$]
    {\includegraphics[scale=0.975]{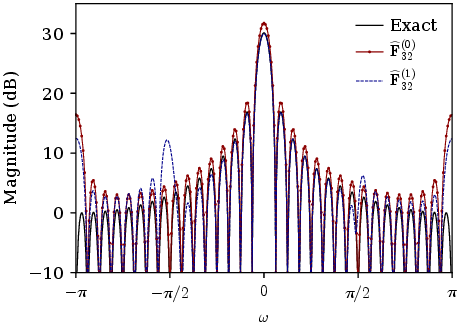}
    \label{fig:resp-v0v1-best}}
    \subfigure
        [Best case: $\angle \hat{F}^{(0)}(\omega; 4)$ and $\angle \hat{F}^{(1)}(\omega; 2)$ ]
        {\includegraphics[scale=0.975]{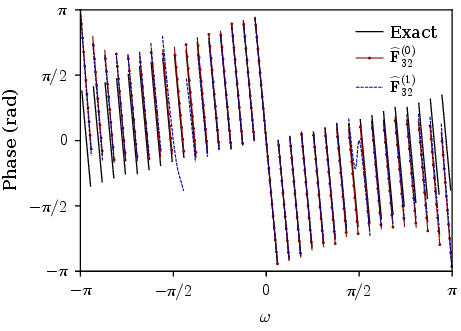}
            \label{fig:phase_T0_T1_best}
        }
    \subfigure[Average case:  $|\hat{F}^{(0)}(\omega; k)|$ and $|\hat{F}^{(1)}(\omega; k)|$]
    {\includegraphics[scale=0.975]{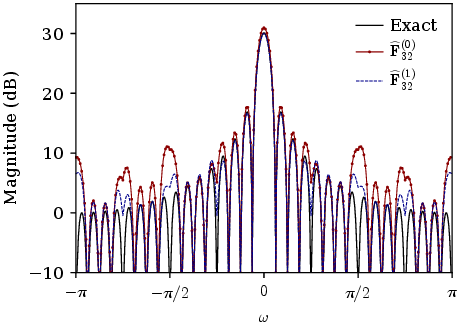}
    \label{fig:resp-v0v1-average}}
        \subfigure
        [Average case: $\angle \hat{F}^{(0)}(\omega; k)$ and
        $\angle \hat{F}^{(1)}(\omega; k)$]
        {\includegraphics[scale=0.975]{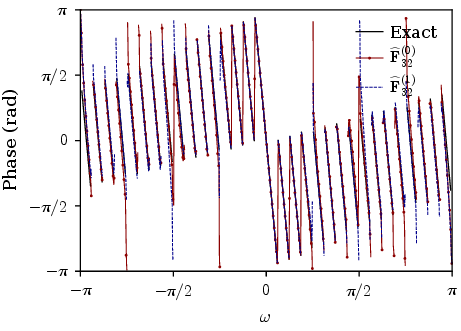}
            \label{fig:phase_T0_T1_average}
        }
    \subfigure[Worst case:  $|\hat{F}^{(0)}(\omega; 1)|$ and $|\hat{F}^{(1)}(\omega; 4)|$]
    {\includegraphics[scale=0.975]{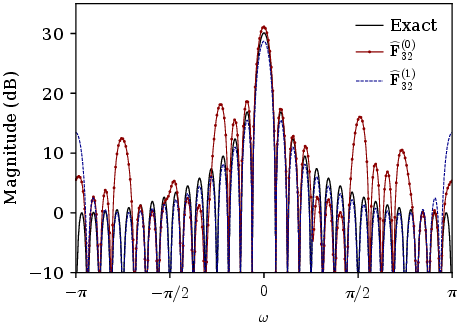}
    \label{fig:resp-v0v1-worst}}
     \subfigure
        [Worst case: $\angle F^{(0)}(\omega; 1)$ and $\angle F^{(1)}(\omega; 4)$ ]
        {\includegraphics[scale=0.975]{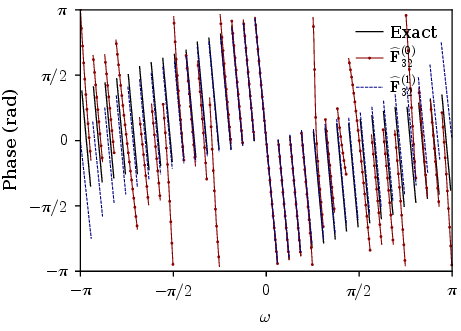}
            \label{fig:phase_T0_T1_worst}
        }
    \caption{Frequency response magnitude
    (left column)
    and phase
    plot comparison:
    exact DFT,
    $\widehat{\mathbf{F}}_{32}^{(0)}$,
    and
    $\widehat{\mathbf{F}}_{32}^{(1)}$
    in
    (a)-(b)~best,
    (c)-(d)~average,
    and
    (e)-(f)~worst cases.%
    }
    \label{fig:frequency-response-selected-cases}
\end{figure}

Figure~\ref{fig:frequency-response-selected-cases}
presents
frequency response
magnitude
and phase plots
associated with
the exact DFT matrix,
the matrix~$\widehat{\mathbf{F}}_{32}^{(0)}$,
and the proposed matrix~$\widehat{\mathbf{F}}_{32}^{(1)}$.
We consider:
(i)~$\hat{F}^{(0)}(\omega;4)$ and $\hat{F}^{(1)}(\omega;2)$ for the best case
(Figure~\ref{fig:frequency-response-selected-cases}(a)-(b));
(ii)~$\hat{F}^{(0)}(\omega;1)$ and $\hat{F}^{(1)}(\omega;4)$ for the worst case
(Figure~\ref{fig:frequency-response-selected-cases}(e)-(f));
and
(iii)~the average case
(Figure~\ref{fig:frequency-response-selected-cases}(c)-(d)).
The average phase curves
were obtained by taking
the mean of the centered individual
frequency response phase curves.

\subsection{Arithmetic Complexity Analysis}

Tables~\ref{complexity-algorithm-ADFT-0} and~\ref{complexity-algorithm-ADFT-1}
summarize
the arithmetic complexity of the factorizations of
$\widehat{\mathbf{F}}_{32}^{(0)}$
and
$\widehat{\mathbf{F}}_{32}^{(1)}$,
respectively.
The factorization of $\widehat{\mathbf{F}}_{32}^{(0)}$ requires
144~real additions
and
no bit-shifts~\cite{cintra2024approximation};
whereas the factorization of $\widehat{\mathbf{F}}_{32}^{(1)}$
requires
152~real additions
and
34~bit-shifts.
The bit-shift operations
are
concentrated in
the factors
$\mathbf{D}_1$,
$\mathbf{D}_2$,
$\mathbf{D}_3$,
$\mathbf{W}_9$,
$\mathbf{W}_{10}$,
and
$\mathbf{W}_{13}$.

\begin{table}
\centering
\caption{%
Arithmetic Complexity Assessment
of $\widehat{\mathbf{F}}_{32}^{(0)}$
Factorization}
\label{complexity-algorithm-ADFT-0}
\begin{tabular}{l cccc cccc | c }
\toprule
Operation &
$\mathbf{W}_1$ &
$\mathbf{W}_2$ &
$\mathbf{W}_3$ &
$\mathbf{W}_4$ &
$\mathbf{W}_5$ &
$\mathbf{W}_6$ &
$\mathbf{W}_7$ &
$\mathbf{W}_8$ &
Total
\\
\midrule
Real multiplications &
0 & 0 & 0 & 0 & 0 & 0 & 0 & 0 & 0
\\
Real additions &
30 & 30 & 14 & 14 & 30 & 14 & 12 & 0 & 144
\\
\bottomrule
\end{tabular}%
\end{table}

\begin{table}
\centering
\caption{%
Arithmetic Complexity Assessment
of $\widehat{\mathbf{F}}_{32}^{(1)}$
Factorization}
\label{complexity-algorithm-ADFT-1}
\begin{tabular}{l cccc cccc ccc | c}
\toprule
Operation &
$\mathbf{W}_1$ &
$\mathbf{W}_2$ &
$\mathbf{W}_3$ &
$\mathbf{D}_1$ &
$\mathbf{W}_9$ &
$\mathbf{D}_2$ &
$\mathbf{W}_{10}$ &
$\mathbf{W}_{11}$ &
$\mathbf{W}_{12}$ &
$\mathbf{D}_3$ &
$\mathbf{W}_{13}$ &
Total
\\
\midrule
Real multiplications &
0 & 0 & 0 & 0 & 0 & 0 & 0 & 0 & 0 & 0 & 0 & 0
\\
Real additions &
30 & 30 & 14 & 0 & 34 & 0 & 22 & 14 & 8 & 0 & 0 & 152
\\
Bit-shifts &
0 & 0 & 0 & 3 & 22 & 3 & 2 & 0 & 0 & 2 & 2 & 34
\\
\bottomrule
\end{tabular}%
\end{table}

\section{Discussion}

In this discussion,
matrix rows~0, 8, 16, and 24
are not considered
because they already
consist of trivial multipliers
in the exact DFT,
resulting in null approximation
error by definition.
In
Tables~\ref{table-energy-error-matrix}
and~\ref{table-energy-error-by-row},
the energy error measurements
for
the proposed normalized approximation~$\hat{\mathbf{F}}_{32}^{(2)}$
seem much lower than the other approximations.
However,
this is simply a measuring artifact
due to the fact that the unitary DFT includes
a scaling factor of~$\frac{1}{\sqrt{N}}$.
Thus,
the energy error measurements
for the normalized case
are identical to the measurements
for
the standard approximation~$\hat{\mathbf{F}}_{32}^{(3)}$,
only
scaled by $\frac{1}{32}$.
Therefore,
in the following,
we remove from our discussion the
measurements associated with
$\hat{\mathbf{F}}_{32}^{(2)}$;
they were provided for the sake of completeness.

\paragraph{Error Measurements.}

Table~\ref{table-energy-error-matrix}
shows that
the proposed raw approximation~$\hat{\mathbf{F}}_{32}^{(1)}$
could offer
better energy-based error measurements
than
the baseline approximation~$\hat{\mathbf{F}}_{32}^{(0)}$~\cite{cintra2024approximation}.
The proposed approximation $\widehat{\mathbf{F}}_{32}^{(1)}$
offers reduced error
when compared to
$\widehat{\mathbf{F}}_{32}^{(0)}$---a
reduction of $71.2\%$ in total energy error and
of $47.3\%$ in MAPE.

In terms of orthogonality,
although not strictly orthogonal,
we emphasize
that the proposed approximation
possesses
a small deviation from orthogonality;
essentially the same deviation
as the one from the baseline method.
Such small deviation from orthogonality renders the proposed approximation
a candidate method for application in areas where orthogonality is
an often sought property,
such as
communications systems~\cite{weinstein1971data, jaradat2019modulation}.
We are not aware of a high-performance,
orthogonal approximation
for the 32-point DFT in the literature.

In terms of the row-wise error analysis,
as shown in Table~\ref{table-energy-error-by-row},
when compared to the exact DFT,
the proposed raw approximation~$\hat{\mathbf{F}}_{32}^{(1)}$
outperforms
the baseline approximation~$\hat{\mathbf{F}}_{32}^{(0)}$,
exhibiting
smaller energy error
for all rows.
Additionally,
the raw approximation
$\hat{\mathbf{F}}_{32}^{(1)}$
offered very close performance
when compared to
the standard DFT approximation
$\hat{\mathbf{F}}_{32}^{(3)}$.
The comparison with
$\hat{\mathbf{F}}_{32}^{(3)}$
is specially
relevant
because
it suggests that
the
matrix term
$
\sqrt{
    \left[
        \operatorname{diag}
            (
            \mathbf{T}
            \cdot
            \mathbf{T}^H
            )
    \right]^{-1}
    }
$
resulting
from the polar decomposition
(cf.~\eqref{equation_polar_definition})
could be suppressed,
depending on the accepted error floor
in the application at hand.
Therefore,
the raw approximation~$\hat{\mathbf{F}}_{32}^{(1)}$
is expected to be
a natural candidate
as
a fundamental core block
for
low-complexity
multiplierless
statistical estimation methods
and
low-cost, low-power
hardware implementations
dedicated to spectral estimation.

\paragraph{Frequency Response.}
We note that rows
of
$\widehat{\mathbf{F}}_{32}^{(0)}$
or
$\widehat{\mathbf{F}}_{32}^{(1)}$
with the same energy error
(cf.~Table~\ref{table-energy-error-by-row})
lead to filters
with identical frequency responses,
up to
a translation (frequency shift)
and/or
a reflection about the vertical axis
(frequency reversal).

Corroborating with the measurements in
Table~\ref{table-energy-error-by-row},
the frequency response magnitude plots
shown in Figure~\ref{fig:frequency-response-comparison}
reveal
that the proposed approximations
incur in
lower spectral leakage
since the sidelobes
in the frequency response magnitude
are consistently closer to
results provided by the exact DFT.
In terms of the phase plots,
when compared to the baseline method,
we notice that the proposed approximation
offers
improved
performance,
exhibiting close agreement
with the exact phase plots.
Again,
this might be compelling evidence
for the application of
the proposed approximations
in the context of
filtering~\cite{ylikaakinen2021frequency},
detection~\cite{wang2024low},
and
communications~\cite{jaradat2019modulation}.

Figure~\ref{fig:frequency-response-selected-cases}
shows that
the proposed approximation
furnishes
superior performance
in terms of phase estimation.
Indeed,
the phase plots from
the proposed
method~$\widehat{\mathbf{F}}_{32}^{(1)}$
display
close agreement with the exact phase,
eliminating or severely reducing the spurious phase disturbances
resulting from
the reference
method~$\widehat{\mathbf{F}}_{32}^{(0)}$.
Such lower phase error
points to the suitability of
the
proposed method
in the context of
phase
estimation~\cite{agrez2005improving, quinn1997estimation}.

\paragraph{Arithmetic Complexity.}
In terms of the arithmetic complexity analysis,
the proposed approximation
admits a fully multiplierless factorization.
Although the proposed approximation operates over a larger numerical domain
$
\left\{
0, \pm \frac{1}{2}, \pm 1
\right\}
$
when compared to
the numerical set
$
\left\{
0, \pm 1
\right\}
$
considered by the reference method,
it could
keep a low arithmetic cost,
very close to that of
the baseline approximation
$\widehat{\mathbf{F}}^{(0)}_{32}$:
only eight extra real additions and
34~bit-shifts.
This modest increase
can be justified
by the favorable results
shown
in the error analysis.
Importantly,
the matrix terms of the
provided factorization
are purely real,
except for the final
factor~$\mathbf{W}_{13}$.
However,
the final factor
does \emph{not} contribute to
any arithmetic cost;
it simply provides
a composition operation
of
the purely real numbers
computed
from the previous matrix factors
to form a complex output
(i.e,
the mapping
$(a,b) \mapsto a + j\cdot b$).
This means
that the proposed approximation
can provide spectral estimation
of real data without requiring complex arithmetic operations.

The trade-off between error and computational cost can be
further analyzed by considering
error-operation measurements.
Table~\ref{table-error-per-op}
shows the discussed error measurements
times the number of additions required by
approximations
$\widehat{\mathbf{F}}_{32}^{(0)}$
and
$\widehat{\mathbf{F}}_{32}^{(1)}$.
Such a combined figure of merit
shows that the proposed method
offers substantial reductions:
\qty{69.63}{\percent}
in energy-error-operation
and
\qty{44.37}{\percent}
in MAPE-operation.

\begin{table}
\centering

\sisetup{
        evaluate-expression = true,
        round-precision   = 1,
        exponent-mode = threshold,
        exponent-thresholds = -2:2,
}

\caption{Error-operation Trade-off
$\widehat{\mathbf{F}}_{32}^{(0)}$
and
$\widehat{\mathbf{F}}_{32}^{(1)}$
}
\label{table-error-per-op}
\begin{tabular}
    {
    l
    S
    S
    }
\toprule
Method &
$\operatorname{\varepsilon} \cdot \text{operation}$ &
$\operatorname{MAPE} \cdot \text{operation}$
\\
\midrule
$
\widehat{\mathbf{F}}_{32}^{(0)}
$~\cite{cintra2024approximation} &
\num{332.230 * 144} & \num{25.913*144}
\\ [8pt]
$\widehat{\mathbf{F}}_{32}^{(1)}$ (proposed) &
\num{95.591 * 152} & \num{13.657*152}
\\
\midrule
\midrule
Variation (\%) &
\qty{(1-(95.591 * 152)/(332.230 * 144))*100}{\percent} &
\qty{(1-(13.657*152)/(25.913*144))*100}{\percent}
\\

\bottomrule
\end{tabular}
\end{table}

\section{Conclusions}

This work introduces a new approximation for the 32-point DFT
as an alternative to
the state-of-the-art design in~\cite{cintra2024approximation}.
The proposed design
is shown to outperform
the reference method,
achieving
a 71.2\% reduction in total energy error and
a 47.3\% reduction in MAPE,
while maintaining a very small
deviation from orthogonality.
Such significantly lower errors
are obtained
at the cost
of \emph{only eight extra additions}.
The frequency response phase plots
indicate that the proposed method
offers better phase estimation properties.
Therefore,
the proposed approximation
is presented as a potentially
favorable option
compared to the design shown
in~\cite{cintra2024approximation},
especially in contexts where
an improved
accuracy-complexity trade-off
is sought.

\section*{Acknowledgments}

The authors thank the Brazilian funding agencies
CNPq (315047/2023-2; 136383/2026-2)
and
CAPES for the partial support.

\appendix

\section{Explicit Numerical Form of the Sparse Factors}
\label{appendix-factors}

The explicit numerical form of
the factorization
of
$
\widehat{\mathbf{F}}_{32}^{(1)}
=
\mathbf{T}^\ast_{32}
$
is provided below:
\begin{align}
\mathbf{W}_1
=
\begin{bsmallmatrix*}[r]
\phantom{-}1 & \phantom{-}0 & \phantom{-}0 & \phantom{-}0 & \phantom{-}0 & \phantom{-}0 & \phantom{-}0 &\phantom{-}0 & \phantom{-}0 & \phantom{-}0 & \phantom{-}0 & \phantom{-}0 & \phantom{-}0 & \phantom{-}0 &\phantom{-}0 & \phantom{-}0 & \phantom{-}1 & \phantom{-}0 & \phantom{-}0 & \phantom{-}0 & \phantom{-}0 &\phantom{-}0 & \phantom{-}0 & \phantom{-}0 & \phantom{-}0 & \phantom{-}0 & \phantom{-}0 & \phantom{-}0 &\phantom{-}0 & \phantom{-}0 & \phantom{-}0 & \phantom{-}0 \\
0 & 1 & 0 & 0 & 0 & 0 & 0 & 0 & 0 & 0 & 0 & 0 & 0 & 0 & 0 & 1 & 0 & 0 & 0 & 0 & 0 & 0 & 0 & 0 & 0 & 0 & 0 & 0 & 0 & 0 & 0 & 0 \\
0 & 0 & 1 & 0 & 0 & 0 & 0 & 0 & 0 & 0 & 0 & 0 & 0 & 0 & 1 & 0 & 0 & 0 & 0 & 0 & 0 & 0 & 0 & 0 & 0 & 0 & 0 & 0 & 0 & 0 & 0 & 0 \\
0 & 0 & 0 & 1 & 0 & 0 & 0 & 0 & 0 & 0 & 0 & 0 & 0 & 1 & 0 & 0 & 0 & 0 & 0 & 0 & 0 & 0 & 0 & 0 & 0 & 0 & 0 & 0 & 0 & 0 & 0 & 0 \\
0 & 0 & 0 & 0 & 1 & 0 & 0 & 0 & 0 & 0 & 0 & 0 & 1 & 0 & 0 & 0 & 0 & 0 & 0 & 0 & 0 & 0 & 0 & 0 & 0 & 0 & 0 & 0 & 0 & 0 & 0 & 0 \\
0 & 0 & 0 & 0 & 0 & 1 & 0 & 0 & 0 & 0 & 0 & 1 & 0 & 0 & 0 & 0 & 0 & 0 & 0 & 0 & 0 & 0 & 0 & 0 & 0 & 0 & 0 & 0 & 0 & 0 & 0 & 0 \\
0 & 0 & 0 & 0 & 0 & 0 & 1 & 0 & 0 & 0 & 1 & 0 & 0 & 0 & 0 & 0 & 0 & 0 & 0 & 0 & 0 & 0 & 0 & 0 & 0 & 0 & 0 & 0 & 0 & 0 & 0 & 0 \\
0 & 0 & 0 & 0 & 0 & 0 & 0 & 1 & 0 & 1 & 0 & 0 & 0 & 0 & 0 & 0 & 0 & 0 & 0 & 0 & 0 & 0 & 0 & 0 & 0 & 0 & 0 & 0 & 0 & 0 & 0 & 0 \\
0 & 0 & 0 & 0 & 0 & 0 & 0 & 0 & 1 & 0 & 0 & 0 & 0 & 0 & 0 & 0 & 0 & 0 & 0 & 0 & 0 & 0 & 0 & 0 & 0 & 0 & 0 & 0 & 0 & 0 & 0 & 0 \\
0 & 0 & 0 & 0 & 0 & 0 & 0 & 1 & 0 & -1 & 0 & 0 & 0 & 0 & 0 & 0 & 0 & 0 & 0 & 0 & 0 & 0 & 0 & 0 & 0 & 0 &0 & 0 & 0 & 0 & 0 & 0 \\
0 & 0 & 0 & 0 & 0 & 0 & 1 & 0 & 0 & 0 & -1 & 0 & 0 & 0 & 0 & 0 & 0 & 0 & 0 & 0 & 0 & 0 & 0 & 0 & 0 & 0 &0 & 0 & 0 & 0 & 0 & 0 \\
0 & 0 & 0 & 0 & 0 & 1 & 0 & 0 & 0 & 0 & 0 & -1 & 0 & 0 & 0 & 0 & 0 & 0 & 0 & 0 & 0 & 0 & 0 & 0 & 0 & 0 &0 & 0 & 0 & 0 & 0 & 0 \\
0 & 0 & 0 & 0 & 1 & 0 & 0 & 0 & 0 & 0 & 0 & 0 & -1 & 0 & 0 & 0 & 0 & 0 & 0 & 0 & 0 & 0 & 0 & 0 & 0 & 0 &0 & 0 & 0 & 0 & 0 & 0 \\
0 & 0 & 0 & 1 & 0 & 0 & 0 & 0 & 0 & 0 & 0 & 0 & 0 & -1 & 0 & 0 & 0 & 0 & 0 & 0 & 0 & 0 & 0 & 0 & 0 & 0 &0 & 0 & 0 & 0 & 0 & 0 \\
0 & 0 & 1 & 0 & 0 & 0 & 0 & 0 & 0 & 0 & 0 & 0 & 0 & 0 & -1 & 0 & 0 & 0 & 0 & 0 & 0 & 0 & 0 & 0 & 0 & 0 &0 & 0 & 0 & 0 & 0 & 0 \\
0 & 1 & 0 & 0 & 0 & 0 & 0 & 0 & 0 & 0 & 0 & 0 & 0 & 0 & 0 & -1 & 0 & 0 & 0 & 0 & 0 & 0 & 0 & 0 & 0 & 0 &0 & 0 & 0 & 0 & 0 & 0 \\
1 & 0 & 0 & 0 & 0 & 0 & 0 & 0 & 0 & 0 & 0 & 0 & 0 & 0 & 0 & 0 & -1 & 0 & 0 & 0 & 0 & 0 & 0 & 0 & 0 & 0 &0 & 0 & 0 & 0 & 0 & 0 \\
0 & 0 & 0 & 0 & 0 & 0 & 0 & 0 & 0 & 0 & 0 & 0 & 0 & 0 & 0 & 0 & 0 & 1 & 0 & 0 & 0 & 0 & 0 & 0 & 0 & 0 & 0 & 0 & 0 & 0 & 0 & 1 \\
0 & 0 & 0 & 0 & 0 & 0 & 0 & 0 & 0 & 0 & 0 & 0 & 0 & 0 & 0 & 0 & 0 & 0 & 1 & 0 & 0 & 0 & 0 & 0 & 0 & 0 & 0 & 0 & 0 & 0 & 1 & 0 \\
0 & 0 & 0 & 0 & 0 & 0 & 0 & 0 & 0 & 0 & 0 & 0 & 0 & 0 & 0 & 0 & 0 & 0 & 0 & 1 & 0 & 0 & 0 & 0 & 0 & 0 & 0 & 0 & 0 & 1 & 0 & 0 \\
0 & 0 & 0 & 0 & 0 & 0 & 0 & 0 & 0 & 0 & 0 & 0 & 0 & 0 & 0 & 0 & 0 & 0 & 0 & 0 & 1 & 0 & 0 & 0 & 0 & 0 & 0 & 0 & 1 & 0 & 0 & 0 \\
0 & 0 & 0 & 0 & 0 & 0 & 0 & 0 & 0 & 0 & 0 & 0 & 0 & 0 & 0 & 0 & 0 & 0 & 0 & 0 & 0 & 1 & 0 & 0 & 0 & 0 & 0 & 1 & 0 & 0 & 0 & 0 \\
0 & 0 & 0 & 0 & 0 & 0 & 0 & 0 & 0 & 0 & 0 & 0 & 0 & 0 & 0 & 0 & 0 & 0 & 0 & 0 & 0 & 0 & 1 & 0 & 0 & 0 & 1 & 0 & 0 & 0 & 0 & 0 \\
0 & 0 & 0 & 0 & 0 & 0 & 0 & 0 & 0 & 0 & 0 & 0 & 0 & 0 & 0 & 0 & 0 & 0 & 0 & 0 & 0 & 0 & 0 & 1 & 0 & 1 & 0 & 0 & 0 & 0 & 0 & 0 \\
0 & 0 & 0 & 0 & 0 & 0 & 0 & 0 & 0 & 0 & 0 & 0 & 0 & 0 & 0 & 0 & 0 & 0 & 0 & 0 & 0 & 0 & 0 & 0 & 1 & 0 & 0 & 0 & 0 & 0 & 0 & 0 \\
0 & 0 & 0 & 0 & 0 & 0 & 0 & 0 & 0 & 0 & 0 & 0 & 0 & 0 & 0 & 0 & 0 & 0 & 0 & 0 & 0 & 0 & 0 & 1 & 0 & -1 &0 & 0 & 0 & 0 & 0 & 0 \\
0 & 0 & 0 & 0 & 0 & 0 & 0 & 0 & 0 & 0 & 0 & 0 & 0 & 0 & 0 & 0 & 0 & 0 & 0 & 0 & 0 & 0 & 1 & 0 & 0 & 0 & -1 & 0 & 0 & 0 & 0 & 0 \\
0 & 0 & 0 & 0 & 0 & 0 & 0 & 0 & 0 & 0 & 0 & 0 & 0 & 0 & 0 & 0 & 0 & 0 & 0 & 0 & 0 & 1 & 0 & 0 & 0 & 0 & 0 & -1 & 0 & 0 & 0 & 0 \\
0 & 0 & 0 & 0 & 0 & 0 & 0 & 0 & 0 & 0 & 0 & 0 & 0 & 0 & 0 & 0 & 0 & 0 & 0 & 0 & 1 & 0 & 0 & 0 & 0 & 0 & 0 & 0 & -1 & 0 & 0 & 0 \\
0 & 0 & 0 & 0 & 0 & 0 & 0 & 0 & 0 & 0 & 0 & 0 & 0 & 0 & 0 & 0 & 0 & 0 & 0 & 1 & 0 & 0 & 0 & 0 & 0 & 0 & 0 & 0 & 0 & -1 & 0 & 0 \\
0 & 0 & 0 & 0 & 0 & 0 & 0 & 0 & 0 & 0 & 0 & 0 & 0 & 0 & 0 & 0 & 0 & 0 & 1 & 0 & 0 & 0 & 0 & 0 & 0 & 0 & 0 & 0 & 0 & 0 & -1 & 0 \\
0 & 0 & 0 & 0 & 0 & 0 & 0 & 0 & 0 & 0 & 0 & 0 & 0 & 0 & 0 & 0 & 0 & 1 & 0 & 0 & 0 & 0 & 0 & 0 & 0 & 0 & 0 & 0 & 0 & 0 & 0 & -1 \\
\end{bsmallmatrix*}
,
\end{align}

\begin{align}
\mathbf{W}_2
=
\begin{bsmallmatrix*}[r]
\phantom{-}1 & \phantom{-}0 & \phantom{-}0 & \phantom{-}0 & \phantom{-}0 & \phantom{-}0 & \phantom{-}0 &\phantom{-}0 & \phantom{-}0 & \phantom{-}0 & \phantom{-}0 & \phantom{-}0 & \phantom{-}0 & \phantom{-}0 &\phantom{-}0 & \phantom{-}0 & \phantom{-}0 & \phantom{-}0 & \phantom{-}0 & \phantom{-}0 & \phantom{-}0 &\phantom{-}0 & \phantom{-}0 & \phantom{-}0 & \phantom{-}0 & \phantom{-}0 & \phantom{-}0 & \phantom{-}0 &\phantom{-}0 & \phantom{-}0 & \phantom{-}0 & \phantom{-}0 \\
0 & 1 & 0 & 0 & 0 & 0 & 0 & 0 & 0 & 0 & 0 & 0 & 0 & 0 & 0 & 0 & 0 & 1 & 0 & 0 & 0 & 0 & 0 & 0 & 0 & 0 & 0 & 0 & 0 & 0 & 0 & 0 \\
0 & 0 & 1 & 0 & 0 & 0 & 0 & 0 & 0 & 0 & 0 & 0 & 0 & 0 & 0 & 0 & 0 & 0 & 1 & 0 & 0 & 0 & 0 & 0 & 0 & 0 & 0 & 0 & 0 & 0 & 0 & 0 \\
0 & 0 & 0 & 1 & 0 & 0 & 0 & 0 & 0 & 0 & 0 & 0 & 0 & 0 & 0 & 0 & 0 & 0 & 0 & 1 & 0 & 0 & 0 & 0 & 0 & 0 & 0 & 0 & 0 & 0 & 0 & 0 \\
0 & 0 & 0 & 0 & 1 & 0 & 0 & 0 & 0 & 0 & 0 & 0 & 0 & 0 & 0 & 0 & 0 & 0 & 0 & 0 & 1 & 0 & 0 & 0 & 0 & 0 & 0 & 0 & 0 & 0 & 0 & 0 \\
0 & 0 & 0 & 0 & 0 & 1 & 0 & 0 & 0 & 0 & 0 & 0 & 0 & 0 & 0 & 0 & 0 & 0 & 0 & 0 & 0 & 1 & 0 & 0 & 0 & 0 & 0 & 0 & 0 & 0 & 0 & 0 \\
0 & 0 & 0 & 0 & 0 & 0 & 1 & 0 & 0 & 0 & 0 & 0 & 0 & 0 & 0 & 0 & 0 & 0 & 0 & 0 & 0 & 0 & 1 & 0 & 0 & 0 & 0 & 0 & 0 & 0 & 0 & 0 \\
0 & 0 & 0 & 0 & 0 & 0 & 0 & 1 & 0 & 0 & 0 & 0 & 0 & 0 & 0 & 0 & 0 & 0 & 0 & 0 & 0 & 0 & 0 & 1 & 0 & 0 & 0 & 0 & 0 & 0 & 0 & 0 \\
0 & 0 & 0 & 0 & 0 & 0 & 0 & 0 & 1 & 0 & 0 & 0 & 0 & 0 & 0 & 0 & 0 & 0 & 0 & 0 & 0 & 0 & 0 & 0 & 1 & 0 & 0 & 0 & 0 & 0 & 0 & 0 \\
0 & 0 & 0 & 0 & 0 & 0 & 0 & 0 & 0 & 1 & 0 & 0 & 0 & 0 & 0 & 0 & 0 & 0 & 0 & 0 & 0 & 0 & 0 & 0 & 0 & 1 & 0 & 0 & 0 & 0 & 0 & 0 \\
0 & 0 & 0 & 0 & 0 & 0 & 0 & 0 & 0 & 0 & 1 & 0 & 0 & 0 & 0 & 0 & 0 & 0 & 0 & 0 & 0 & 0 & 0 & 0 & 0 & 0 & 1 & 0 & 0 & 0 & 0 & 0 \\
0 & 0 & 0 & 0 & 0 & 0 & 0 & 0 & 0 & 0 & 0 & 1 & 0 & 0 & 0 & 0 & 0 & 0 & 0 & 0 & 0 & 0 & 0 & 0 & 0 & 0 & 0 & 1 & 0 & 0 & 0 & 0 \\
0 & 0 & 0 & 0 & 0 & 0 & 0 & 0 & 0 & 0 & 0 & 0 & 1 & 0 & 0 & 0 & 0 & 0 & 0 & 0 & 0 & 0 & 0 & 0 & 0 & 0 & 0 & 0 & 1 & 0 & 0 & 0 \\
0 & 0 & 0 & 0 & 0 & 0 & 0 & 0 & 0 & 0 & 0 & 0 & 0 & 1 & 0 & 0 & 0 & 0 & 0 & 0 & 0 & 0 & 0 & 0 & 0 & 0 & 0 & 0 & 0 & 1 & 0 & 0 \\
0 & 0 & 0 & 0 & 0 & 0 & 0 & 0 & 0 & 0 & 0 & 0 & 0 & 0 & 1 & 0 & 0 & 0 & 0 & 0 & 0 & 0 & 0 & 0 & 0 & 0 & 0 & 0 & 0 & 0 & 1 & 0 \\
0 & 0 & 0 & 0 & 0 & 0 & 0 & 0 & 0 & 0 & 0 & 0 & 0 & 0 & 0 & 1 & 0 & 0 & 0 & 0 & 0 & 0 & 0 & 0 & 0 & 0 & 0 & 0 & 0 & 0 & 0 & 1 \\
0 & 0 & 0 & 0 & 0 & 0 & 0 & 0 & 0 & 0 & 0 & 0 & 0 & 0 & 0 & 0 & 1 & 0 & 0 & 0 & 0 & 0 & 0 & 0 & 0 & 0 & 0 & 0 & 0 & 0 & 0 & 0 \\
0 & 1 & 0 & 0 & 0 & 0 & 0 & 0 & 0 & 0 & 0 & 0 & 0 & 0 & 0 & 0 & 0 & -1 & 0 & 0 & 0 & 0 & 0 & 0 & 0 & 0 &0 & 0 & 0 & 0 & 0 & 0 \\
0 & 0 & 1 & 0 & 0 & 0 & 0 & 0 & 0 & 0 & 0 & 0 & 0 & 0 & 0 & 0 & 0 & 0 & -1 & 0 & 0 & 0 & 0 & 0 & 0 & 0 &0 & 0 & 0 & 0 & 0 & 0 \\
0 & 0 & 0 & 1 & 0 & 0 & 0 & 0 & 0 & 0 & 0 & 0 & 0 & 0 & 0 & 0 & 0 & 0 & 0 & -1 & 0 & 0 & 0 & 0 & 0 & 0 &0 & 0 & 0 & 0 & 0 & 0 \\
0 & 0 & 0 & 0 & 1 & 0 & 0 & 0 & 0 & 0 & 0 & 0 & 0 & 0 & 0 & 0 & 0 & 0 & 0 & 0 & -1 & 0 & 0 & 0 & 0 & 0 &0 & 0 & 0 & 0 & 0 & 0 \\
0 & 0 & 0 & 0 & 0 & 1 & 0 & 0 & 0 & 0 & 0 & 0 & 0 & 0 & 0 & 0 & 0 & 0 & 0 & 0 & 0 & -1 & 0 & 0 & 0 & 0 &0 & 0 & 0 & 0 & 0 & 0 \\
0 & 0 & 0 & 0 & 0 & 0 & 1 & 0 & 0 & 0 & 0 & 0 & 0 & 0 & 0 & 0 & 0 & 0 & 0 & 0 & 0 & 0 & -1 & 0 & 0 & 0 &0 & 0 & 0 & 0 & 0 & 0 \\
0 & 0 & 0 & 0 & 0 & 0 & 0 & 1 & 0 & 0 & 0 & 0 & 0 & 0 & 0 & 0 & 0 & 0 & 0 & 0 & 0 & 0 & 0 & -1 & 0 & 0 &0 & 0 & 0 & 0 & 0 & 0 \\
0 & 0 & 0 & 0 & 0 & 0 & 0 & 0 & 1 & 0 & 0 & 0 & 0 & 0 & 0 & 0 & 0 & 0 & 0 & 0 & 0 & 0 & 0 & 0 & -1 & 0 &0 & 0 & 0 & 0 & 0 & 0 \\
0 & 0 & 0 & 0 & 0 & 0 & 0 & 0 & 0 & 1 & 0 & 0 & 0 & 0 & 0 & 0 & 0 & 0 & 0 & 0 & 0 & 0 & 0 & 0 & 0 & -1 &0 & 0 & 0 & 0 & 0 & 0 \\
0 & 0 & 0 & 0 & 0 & 0 & 0 & 0 & 0 & 0 & 1 & 0 & 0 & 0 & 0 & 0 & 0 & 0 & 0 & 0 & 0 & 0 & 0 & 0 & 0 & 0 & -1 & 0 & 0 & 0 & 0 & 0 \\
0 & 0 & 0 & 0 & 0 & 0 & 0 & 0 & 0 & 0 & 0 & 1 & 0 & 0 & 0 & 0 & 0 & 0 & 0 & 0 & 0 & 0 & 0 & 0 & 0 & 0 & 0 & -1 & 0 & 0 & 0 & 0 \\
0 & 0 & 0 & 0 & 0 & 0 & 0 & 0 & 0 & 0 & 0 & 0 & 1 & 0 & 0 & 0 & 0 & 0 & 0 & 0 & 0 & 0 & 0 & 0 & 0 & 0 & 0 & 0 & -1 & 0 & 0 & 0 \\
0 & 0 & 0 & 0 & 0 & 0 & 0 & 0 & 0 & 0 & 0 & 0 & 0 & 1 & 0 & 0 & 0 & 0 & 0 & 0 & 0 & 0 & 0 & 0 & 0 & 0 & 0 & 0 & 0 & -1 & 0 & 0 \\
0 & 0 & 0 & 0 & 0 & 0 & 0 & 0 & 0 & 0 & 0 & 0 & 0 & 0 & 1 & 0 & 0 & 0 & 0 & 0 & 0 & 0 & 0 & 0 & 0 & 0 & 0 & 0 & 0 & 0 & -1 & 0 \\
0 & 0 & 0 & 0 & 0 & 0 & 0 & 0 & 0 & 0 & 0 & 0 & 0 & 0 & 0 & 1 & 0 & 0 & 0 & 0 & 0 & 0 & 0 & 0 & 0 & 0 & 0 & 0 & 0 & 0 & 0 & -1 \\
\end{bsmallmatrix*}
,
\end{align}

\begin{align}
\mathbf{W}_3
=
\begin{bsmallmatrix*}[r]
\phantom{-}1 & \phantom{-}0 & \phantom{-}0 & \phantom{-}0 & \phantom{-}0 & \phantom{-}0 & \phantom{-}0 &\phantom{-}0 & \phantom{-}1 & \phantom{-}0 & \phantom{-}0 & \phantom{-}0 & \phantom{-}0 & \phantom{-}0 &\phantom{-}0 & \phantom{-}0 & \phantom{-}0 & \phantom{-}0 & \phantom{-}0 & \phantom{-}0 & \phantom{-}0 &\phantom{-}0 & \phantom{-}0 & \phantom{-}0 & \phantom{-}0 & \phantom{-}0 & \phantom{-}0 & \phantom{-}0 &\phantom{-}0 & \phantom{-}0 & \phantom{-}0 & \phantom{-}0 \\
0 & 1 & 0 & 0 & 0 & 0 & 0 & 1 & 0 & 0 & 0 & 0 & 0 & 0 & 0 & 0 & 0 & 0 & 0 & 0 & 0 & 0 & 0 & 0 & 0 & 0 & 0 & 0 & 0 & 0 & 0 & 0 \\
0 & 0 & 1 & 0 & 0 & 0 & 1 & 0 & 0 & 0 & 0 & 0 & 0 & 0 & 0 & 0 & 0 & 0 & 0 & 0 & 0 & 0 & 0 & 0 & 0 & 0 & 0 & 0 & 0 & 0 & 0 & 0 \\
0 & 0 & 0 & 1 & 0 & 1 & 0 & 0 & 0 & 0 & 0 & 0 & 0 & 0 & 0 & 0 & 0 & 0 & 0 & 0 & 0 & 0 & 0 & 0 & 0 & 0 & 0 & 0 & 0 & 0 & 0 & 0 \\
0 & 0 & 0 & 0 & 1 & 0 & 0 & 0 & 0 & 0 & 0 & 0 & 0 & 0 & 0 & 0 & 0 & 0 & 0 & 0 & 0 & 0 & 0 & 0 & 0 & 0 & 0 & 0 & 0 & 0 & 0 & 0 \\
0 & 0 & 0 & 1 & 0 & -1 & 0 & 0 & 0 & 0 & 0 & 0 & 0 & 0 & 0 & 0 & 0 & 0 & 0 & 0 & 0 & 0 & 0 & 0 & 0 & 0 &0 & 0 & 0 & 0 & 0 & 0 \\
0 & 0 & 1 & 0 & 0 & 0 & -1 & 0 & 0 & 0 & 0 & 0 & 0 & 0 & 0 & 0 & 0 & 0 & 0 & 0 & 0 & 0 & 0 & 0 & 0 & 0 &0 & 0 & 0 & 0 & 0 & 0 \\
0 & 1 & 0 & 0 & 0 & 0 & 0 & -1 & 0 & 0 & 0 & 0 & 0 & 0 & 0 & 0 & 0 & 0 & 0 & 0 & 0 & 0 & 0 & 0 & 0 & 0 &0 & 0 & 0 & 0 & 0 & 0 \\
1 & 0 & 0 & 0 & 0 & 0 & 0 & 0 & -1 & 0 & 0 & 0 & 0 & 0 & 0 & 0 & 0 & 0 & 0 & 0 & 0 & 0 & 0 & 0 & 0 & 0 &0 & 0 & 0 & 0 & 0 & 0 \\
0 & 0 & 0 & 0 & 0 & 0 & 0 & 0 & 0 & 1 & 0 & 0 & 0 & 0 & 0 & 1 & 0 & 0 & 0 & 0 & 0 & 0 & 0 & 0 & 0 & 0 & 0 & 0 & 0 & 0 & 0 & 0 \\
0 & 0 & 0 & 0 & 0 & 0 & 0 & 0 & 0 & 0 & 1 & 0 & 0 & 0 & 1 & 0 & 0 & 0 & 0 & 0 & 0 & 0 & 0 & 0 & 0 & 0 & 0 & 0 & 0 & 0 & 0 & 0 \\
0 & 0 & 0 & 0 & 0 & 0 & 0 & 0 & 0 & 0 & 0 & 1 & 0 & 1 & 0 & 0 & 0 & 0 & 0 & 0 & 0 & 0 & 0 & 0 & 0 & 0 & 0 & 0 & 0 & 0 & 0 & 0 \\
0 & 0 & 0 & 0 & 0 & 0 & 0 & 0 & 0 & 0 & 0 & 0 & 1 & 0 & 0 & 0 & 0 & 0 & 0 & 0 & 0 & 0 & 0 & 0 & 0 & 0 & 0 & 0 & 0 & 0 & 0 & 0 \\
0 & 0 & 0 & 0 & 0 & 0 & 0 & 0 & 0 & 0 & 0 & 1 & 0 & -1 & 0 & 0 & 0 & 0 & 0 & 0 & 0 & 0 & 0 & 0 & 0 & 0 &0 & 0 & 0 & 0 & 0 & 0 \\
0 & 0 & 0 & 0 & 0 & 0 & 0 & 0 & 0 & 0 & 1 & 0 & 0 & 0 & -1 & 0 & 0 & 0 & 0 & 0 & 0 & 0 & 0 & 0 & 0 & 0 &0 & 0 & 0 & 0 & 0 & 0 \\
0 & 0 & 0 & 0 & 0 & 0 & 0 & 0 & 0 & 1 & 0 & 0 & 0 & 0 & 0 & -1 & 0 & 0 & 0 & 0 & 0 & 0 & 0 & 0 & 0 & 0 &0 & 0 & 0 & 0 & 0 & 0 \\
0 & 0 & 0 & 0 & 0 & 0 & 0 & 0 & 0 & 0 & 0 & 0 & 0 & 0 & 0 & 0 & 1 & 0 & 0 & 0 & 0 & 0 & 0 & 0 & 0 & 0 & 0 & 0 & 0 & 0 & 0 & 0 \\
0 & 0 & 0 & 0 & 0 & 0 & 0 & 0 & 0 & 0 & 0 & 0 & 0 & 0 & 0 & 0 & 0 & 1 & 0 & 0 & 0 & 0 & 0 & 0 & 0 & 0 & 0 & 0 & 0 & 0 & 0 & 0 \\
0 & 0 & 0 & 0 & 0 & 0 & 0 & 0 & 0 & 0 & 0 & 0 & 0 & 0 & 0 & 0 & 0 & 0 & 1 & 0 & 0 & 0 & 0 & 0 & 0 & 0 & 0 & 0 & 0 & 0 & 0 & 0 \\
0 & 0 & 0 & 0 & 0 & 0 & 0 & 0 & 0 & 0 & 0 & 0 & 0 & 0 & 0 & 0 & 0 & 0 & 0 & 1 & 0 & 0 & 0 & 0 & 0 & 0 & 0 & 0 & 0 & 0 & 0 & 0 \\
0 & 0 & 0 & 0 & 0 & 0 & 0 & 0 & 0 & 0 & 0 & 0 & 0 & 0 & 0 & 0 & 0 & 0 & 0 & 0 & 1 & 0 & 0 & 0 & 0 & 0 & 0 & 0 & 0 & 0 & 0 & 0 \\
0 & 0 & 0 & 0 & 0 & 0 & 0 & 0 & 0 & 0 & 0 & 0 & 0 & 0 & 0 & 0 & 0 & 0 & 0 & 0 & 0 & 1 & 0 & 0 & 0 & 0 & 0 & 0 & 0 & 0 & 0 & 0 \\
0 & 0 & 0 & 0 & 0 & 0 & 0 & 0 & 0 & 0 & 0 & 0 & 0 & 0 & 0 & 0 & 0 & 0 & 0 & 0 & 0 & 0 & 1 & 0 & 0 & 0 & 0 & 0 & 0 & 0 & 0 & 0 \\
0 & 0 & 0 & 0 & 0 & 0 & 0 & 0 & 0 & 0 & 0 & 0 & 0 & 0 & 0 & 0 & 0 & 0 & 0 & 0 & 0 & 0 & 0 & 1 & 0 & 0 & 0 & 0 & 0 & 0 & 0 & 0 \\
0 & 0 & 0 & 0 & 0 & 0 & 0 & 0 & 0 & 0 & 0 & 0 & 0 & 0 & 0 & 0 & 0 & 0 & 0 & 0 & 0 & 0 & 0 & 0 & 1 & 0 & 0 & 0 & 0 & 0 & 0 & 0 \\
0 & 0 & 0 & 0 & 0 & 0 & 0 & 0 & 0 & 0 & 0 & 0 & 0 & 0 & 0 & 0 & 0 & 0 & 0 & 0 & 0 & 0 & 0 & 0 & 0 & 1 & 0 & 0 & 0 & 0 & 0 & 0 \\
0 & 0 & 0 & 0 & 0 & 0 & 0 & 0 & 0 & 0 & 0 & 0 & 0 & 0 & 0 & 0 & 0 & 0 & 0 & 0 & 0 & 0 & 0 & 0 & 0 & 0 & 1 & 0 & 0 & 0 & 0 & 0 \\
0 & 0 & 0 & 0 & 0 & 0 & 0 & 0 & 0 & 0 & 0 & 0 & 0 & 0 & 0 & 0 & 0 & 0 & 0 & 0 & 0 & 0 & 0 & 0 & 0 & 0 & 0 & 1 & 0 & 0 & 0 & 0 \\
0 & 0 & 0 & 0 & 0 & 0 & 0 & 0 & 0 & 0 & 0 & 0 & 0 & 0 & 0 & 0 & 0 & 0 & 0 & 0 & 0 & 0 & 0 & 0 & 0 & 0 & 0 & 0 & 1 & 0 & 0 & 0 \\
0 & 0 & 0 & 0 & 0 & 0 & 0 & 0 & 0 & 0 & 0 & 0 & 0 & 0 & 0 & 0 & 0 & 0 & 0 & 0 & 0 & 0 & 0 & 0 & 0 & 0 & 0 & 0 & 0 & 1 & 0 & 0 \\
0 & 0 & 0 & 0 & 0 & 0 & 0 & 0 & 0 & 0 & 0 & 0 & 0 & 0 & 0 & 0 & 0 & 0 & 0 & 0 & 0 & 0 & 0 & 0 & 0 & 0 & 0 & 0 & 0 & 0 & 1 & 0 \\
0 & 0 & 0 & 0 & 0 & 0 & 0 & 0 & 0 & 0 & 0 & 0 & 0 & 0 & 0 & 0 & 0 & 0 & 0 & 0 & 0 & 0 & 0 & 0 & 0 & 0 & 0 & 0 & 0 & 0 & 0 & 1 \\
\end{bsmallmatrix*}
,
\end{align}

\begin{align}
	\mathbf{W}_9
	=
	\begin{bsmallmatrix*}[r]
		\phantom{-}1 & \phantom{-}0 & \phantom{-}0 & \phantom{-}0 & \phantom{-}1 & \phantom{-}0 & \phantom{-}0 & \phantom{-}0 & \phantom{-}0 & \phantom{-}0 & \phantom{-}0 & \phantom{-}0 & \phantom{-}0 & \phantom{-}0 & \phantom{-}0 & \phantom{-}0 & \phantom{-}0 & \phantom{-}0 & \phantom{-}0 & \phantom{-}0 & \phantom{-}0 & \phantom{-}0 & \phantom{-}0 & \phantom{-}0 & \phantom{-}0 & \phantom{-}0 & \phantom{-}0 & \phantom{-}0 & \phantom{-}0 & \phantom{-}0 & \phantom{-}0 & \phantom{-}0 \\
		0 & 1 & 0 & 1 & 0 & 0 & 0 & 0 & 0 & 0 & 0 & 0 & 0 & 0 & 0 & 0 & 0 & 0 & 0 & 0 & 0 & 0 & 0 & 0 & 0 & 0 & 0 & 0 & 0 & 0 & 0 & 0 \\
		0 & 0 & 1 & 0 & 0 & 0 & 0 & 0 & 0 & 0 & 0 & 0 & 0 & 0 & 0 & 0 & 0 & 0 & 0 & 0 & 0 & 0 & 0 & 0 & 0 & 0 & 0 & 0 & 0 & 0 & 0 & 0 \\
		0 & 1 & 0 & -1 & 0 & 0 & 0 & 0 & 0 & 0 & 0 & 0 & 0 & 0 & 0 & 0 & 0 & 0 & 0 & 0 & 0 & 0 & 0 & 0 & 0 & 0 & 0 & 0 & 0 & 0 & 0 & 0 \\
		1 & 0 & 0 & 0 & -1 & 0 & 0 & 0 & 0 & 0 & 0 & 0 & 0 & 0 & 0 & 0 & 0 & 0 & 0 & 0 & 0 & 0 & 0 & 0 & 0 & 0 & 0 & 0 & 0 & 0 & 0 & 0 \\
		0 & 0 & 0 & 0 & 0 & 1 & 0 & 2 & 0 & 0 & 0 & 0 & 0 & 0 & 0 & 0 & 0 & 0 & 0 & 0 & 0 & 0 & 0 & 0 & 0 & 0 & 0 & 0 & 0 & 0 & 0 & 0 \\
		0 & 0 & 0 & 0 & 0 & 0 & 1 & 0 & 1 & 0 & 0 & 0 & 0 & 0 & 0 & 0 & 0 & 0 & 0 & 0 & 0 & 0 & 0 & 0 & 0 & 0 & 0 & 0 & 0 & 0 & 0 & 0 \\
		0 & 0 & 0 & 0 & 0 & -2 & 0 & 1 & 0 & 0 & 0 & 0 & 0 & 0 & 0 & 0 & 0 & 0 & 0 & 0 & 0 & 0 & 0 & 0 & 0 & 0 & 0 & 0 & 0 & 0 & 0 & 0 \\
		0 & 0 & 0 & 0 & 0 & 0 & -1 & 0 & 1 & 0 & 0 & 0 & 0 & 0 & 0 & 0 & 0 & 0 & 0 & 0 & 0 & 0 & 0 & 0 & 0 & 0 & 0 & 0 & 0 & 0 & 0 & 0 \\
		0 & 0 & 0 & 0 & 0 & 0 & 0 & 0 & 0 & 1 & 0 & 2 & 0 & 0 & 0 & 0 & 0 & 0 & 0 & 0 & 0 & 0 & 0 & 0 & 0 & 0 & 0 & 0 & 0 & 0 & 0 & 0 \\
		0 & 0 & 0 & 0 & 0 & 0 & 0 & 0 & 0 & 0 & 1 & 0 & 1 & 0 & 0 & 0 & 0 & 0 & 0 & 0 & 0 & 0 & 0 & 0 & 0 & 0 & 0 & 0 & 0 & 0 & 0 & 0 \\
		0 & 0 & 0 & 0 & 0 & 0 & 0 & 0 & 0 & -2 & 0 & 1 & 0 & 0 & 0 & 0 & 0 & 0 & 0 & 0 & 0 & 0 & 0 & 0 & 0 & 0 & 0 & 0 & 0 & 0 & 0 & 0 \\
		0 & 0 & 0 & 0 & 0 & 0 & 0 & 0 & 0 & 0 & -1 & 0 & 1 & 0 & 0 & 0 & 0 & 0 & 0 & 0 & 0 & 0 & 0 & 0 & 0 & 0 & 0 & 0 & 0 & 0 & 0 & 0 \\
		0 & 0 & 0 & 0 & 0 & 0 & 0 & 0 & 0 & 0 & 0 & 0 & 0 & 1 & 0 & 1 & 0 & 0 & 0 & 0 & 0 & 0 & 0 & 0 & 0 & 0 & 0 & 0 & 0 & 0 & 0 & 0 \\
		0 & 0 & 0 & 0 & 0 & 0 & 0 & 0 & 0 & 0 & 0 & 0 & 0 & 0 & 1 & 0 & 0 & 0 & 0 & 0 & 0 & 0 & 0 & 0 & 0 & 0 & 0 & 0 & 0 & 0 & 0 & 0 \\
		0 & 0 & 0 & 0 & 0 & 0 & 0 & 0 & 0 & 0 & 0 & 0 & 0 & 1 & 0 & -1 & 0 & 0 & 0 & 0 & 0 & 0 & 0 & 0 & 0 & 0 & 0 & 0 & 0 & 0 & 0 & 0 \\
		0 & 0 & 0 & 0 & 0 & 0 & 0 & 0 & 0 & 0 & 0 & 0 & 0 & 0 & 0 & 0 & 1 & 0 & 0 & 0 & 0 & 0 & 0 & 0 & 0 & 0 & 0 & 0 & 0 & 0 & 0 & 0 \\
		0 & 0 & 0 & 0 & 0 & 0 & 0 & 0 & 0 & 0 & 0 & 0 & 0 & 0 & 0 & 0 & 0 & 0 & 0 & -1 & 0 & -2 & 0 & -2 & 0 & 0 & 0 & 0 & 0 & 0 & 0 & 0 \\
		0 & 0 & 0 & 0 & 0 & 0 & 0 & 0 & 0 & 0 & 0 & 0 & 0 & 0 & 0 & 0 & 0 & 0 & 1 & 0 & 0 & 0 & 2 & 0 & 0 & 0 & 0 & 0 & 0 & 0 & 0 & 0 \\
		0 & 0 & 0 & 0 & 0 & 0 & 0 & 0 & 0 & 0 & 0 & 0 & 0 & 0 & 0 & 0 & 0 & -1 & 0 & -2 & 0 & 0 & 0 & 2 & 0 & 0 & 0 & 0 & 0 & 0 & 0 & 0 \\
		0 & 0 & 0 & 0 & 0 & 0 & 0 & 0 & 0 & 0 & 0 & 0 & 0 & 0 & 0 & 0 & 0 & 0 & 0 & 0 & 1 & 0 & 0 & 0 & 1 & 0 & 0 & 0 & 0 & 0 & 0 & 0 \\
		0 & 0 & 0 & 0 & 0 & 0 & 0 & 0 & 0 & 0 & 0 & 0 & 0 & 0 & 0 & 0 & 0 & -2 & 0 & 0 & 0 & 2 & 0 & -1 & 0 & 0 & 0 & 0 & 0 & 0 & 0 & 0 \\
		0 & 0 & 0 & 0 & 0 & 0 & 0 & 0 & 0 & 0 & 0 & 0 & 0 & 0 & 0 & 0 & 0 & 0 & -2 & 0 & 0 & 0 & 1 & 0 & 0 & 0 & 0 & 0 & 0 & 0 & 0 & 0 \\
		0 & 0 & 0 & 0 & 0 & 0 & 0 & 0 & 0 & 0 & 0 & 0 & 0 & 0 & 0 & 0 & 0 & -2 & 0 & 2 & 0 & -1 & 0 & 0 & 0 & 0 & 0 & 0 & 0 & 0 & 0 & 0 \\
		0 & 0 & 0 & 0 & 0 & 0 & 0 & 0 & 0 & 0 & 0 & 0 & 0 & 0 & 0 & 0 & 0 & 0 & 0 & 0 & -1 & 0 & 0 & 0 & 1 & 0 & 0 & 0 & 0 & 0 & 0 & 0 \\
		0 & 0 & 0 & 0 & 0 & 0 & 0 & 0 & 0 & 0 & 0 & 0 & 0 & 0 & 0 & 0 & 0 & 0 & 0 & 0 & 0 & 0 & 0 & 0 & 0 & 0 & 0 & -1 & 0 & -2 & 0 & -2 \\
		0 & 0 & 0 & 0 & 0 & 0 & 0 & 0 & 0 & 0 & 0 & 0 & 0 & 0 & 0 & 0 & 0 & 0 & 0 & 0 & 0 & 0 & 0 & 0 & 0 & 0 & 1 & 0 & 0 & 0 & 0 & 0 \\
		0 & 0 & 0 & 0 & 0 & 0 & 0 & 0 & 0 & 0 & 0 & 0 & 0 & 0 & 0 & 0 & 0 & 0 & 0 & 0 & 0 & 0 & 0 & 0 & 0 & -1 & 0 & -2 & 0 & 0 & 0 & 2 \\
		0 & 0 & 0 & 0 & 0 & 0 & 0 & 0 & 0 & 0 & 0 & 0 & 0 & 0 & 0 & 0 & 0 & 0 & 0 & 0 & 0 & 0 & 0 & 0 & 0 & 0 & 0 & 0 & 1 & 0 & 0 & 0 \\
		0 & 0 & 0 & 0 & 0 & 0 & 0 & 0 & 0 & 0 & 0 & 0 & 0 & 0 & 0 & 0 & 0 & 0 & 0 & 0 & 0 & 0 & 0 & 0 & 0 & -2 & 0 & 0 & 0 & 2 & 0 & -1 \\
		0 & 0 & 0 & 0 & 0 & 0 & 0 & 0 & 0 & 0 & 0 & 0 & 0 & 0 & 0 & 0 & 0 & 0 & 0 & 0 & 0 & 0 & 0 & 0 & 0 & 0 & 0 & 0 & 0 & 0 & 1 & 0 \\
		0 & 0 & 0 & 0 & 0 & 0 & 0 & 0 & 0 & 0 & 0 & 0 & 0 & 0 & 0 & 0 & 0 & 0 & 0 & 0 & 0 & 0 & 0 & 0 & 0 & -2 & 0 & 2 & 0 & -1 & 0 & 0
	\end{bsmallmatrix*}
	,
\end{align}

\begin{align}
	\mathbf{W}_{10}
	=
	\begin{bsmallmatrix*}[r]
		\phantom{-}1 & \phantom{-}0 & \phantom{-}1 & \phantom{-}0 & \phantom{-}0 & \phantom{-}0 & \phantom{-}0 & \phantom{-}0 & \phantom{-}0 & \phantom{-}0 & \phantom{-}0 & \phantom{-}0 & \phantom{-}0 & \phantom{-}0 & \phantom{-}0 & \phantom{-}0 & \phantom{-}0 & \phantom{-}0 & \phantom{-}0 & \phantom{-}0 & \phantom{-}0 & \phantom{-}0 & \phantom{-}0 & \phantom{-}0 & \phantom{-}0 & \phantom{-}0 & \phantom{-}0 & \phantom{-}0 & \phantom{-}0 & \phantom{-}0 & \phantom{-}0 & \phantom{-}0 \\
		0 & 1 & 0 & 0 & 0 & 0 & 0 & 0 & 0 & 0 & 0 & 0 & 0 & 0 & 0 & 0 & 0 & 0 & 0 & 0 & 0 & 0 & 0 & 0 & 0 & 0 & 0 & 0 & 0 & 0 & 0 & 0 \\
		1 & 0 & -1 & 0 & 0 & 0 & 0 & 0 & 0 & 0 & 0 & 0 & 0 & 0 & 0 & 0 & 0 & 0 & 0 & 0 & 0 & 0 & 0 & 0 & 0 & 0 & 0 & 0 & 0 & 0 & 0 & 0 \\
		0 & 0 & 0 & 1 & 1 & 0 & 0 & 0 & 0 & 0 & 0 & 0 & 0 & 0 & 0 & 0 & 0 & 0 & 0 & 0 & 0 & 0 & 0 & 0 & 0 & 0 & 0 & 0 & 0 & 0 & 0 & 0 \\
		0 & 0 & 0 & -1 & 1 & 0 & 0 & 0 & 0 & 0 & 0 & 0 & 0 & 0 & 0 & 0 & 0 & 0 & 0 & 0 & 0 & 0 & 0 & 0 & 0 & 0 & 0 & 0 & 0 & 0 & 0 & 0 \\
		0 & 0 & 0 & 0 & 0 & 1 & 1 & 0 & 0 & 0 & 0 & 0 & 0 & 0 & 0 & 0 & 0 & 0 & 0 & 0 & 0 & 0 & 0 & 0 & 0 & 0 & 0 & 0 & 0 & 0 & 0 & 0 \\
		0 & 0 & 0 & 0 & 0 & 1 & -1 & 0 & 0 & 0 & 0 & 0 & 0 & 0 & 0 & 0 & 0 & 0 & 0 & 0 & 0 & 0 & 0 & 0 & 0 & 0 & 0 & 0 & 0 & 0 & 0 & 0 \\
		0 & 0 & 0 & 0 & 0 & 0 & 0 & 1 & 1 & 0 & 0 & 0 & 0 & 0 & 0 & 0 & 0 & 0 & 0 & 0 & 0 & 0 & 0 & 0 & 0 & 0 & 0 & 0 & 0 & 0 & 0 & 0 \\
		0 & 0 & 0 & 0 & 0 & 0 & 0 & 1 & -1 & 0 & 0 & 0 & 0 & 0 & 0 & 0 & 0 & 0 & 0 & 0 & 0 & 0 & 0 & 0 & 0 & 0 & 0 & 0 & 0 & 0 & 0 & 0 \\
		0 & 0 & 0 & 0 & 0 & 0 & 0 & 0 & 0 & 1 & 1 & 0 & 0 & 0 & 0 & 0 & 0 & 0 & 0 & 0 & 0 & 0 & 0 & 0 & 0 & 0 & 0 & 0 & 0 & 0 & 0 & 0 \\
		0 & 0 & 0 & 0 & 0 & 0 & 0 & 0 & 0 & 1 & -1 & 0 & 0 & 0 & 0 & 0 & 0 & 0 & 0 & 0 & 0 & 0 & 0 & 0 & 0 & 0 & 0 & 0 & 0 & 0 & 0 & 0 \\
		0 & 0 & 0 & 0 & 0 & 0 & 0 & 0 & 0 & 0 & 0 & 1 & 1 & 0 & 0 & 0 & 0 & 0 & 0 & 0 & 0 & 0 & 0 & 0 & 0 & 0 & 0 & 0 & 0 & 0 & 0 & 0 \\
		0 & 0 & 0 & 0 & 0 & 0 & 0 & 0 & 0 & 0 & 0 & 1 & -1 & 0 & 0 & 0 & 0 & 0 & 0 & 0 & 0 & 0 & 0 & 0 & 0 & 0 & 0 & 0 & 0 & 0 & 0 & 0 \\
		0 & 0 & 0 & 0 & 0 & 0 & 0 & 0 & 0 & 0 & 0 & 0 & 0 & 1 & 1 & 0 & 0 & 0 & 0 & 0 & 0 & 0 & 0 & 0 & 0 & 0 & 0 & 0 & 0 & 0 & 0 & 0 \\
		0 & 0 & 0 & 0 & 0 & 0 & 0 & 0 & 0 & 0 & 0 & 0 & 0 & -1 & 1 & 0 & 0 & 0 & 0 & 0 & 0 & 0 & 0 & 0 & 0 & 0 & 0 & 0 & 0 & 0 & 0 & 0 \\
		0 & 0 & 0 & 0 & 0 & 0 & 0 & 0 & 0 & 0 & 0 & 0 & 0 & 0 & 0 & 1 & 0 & 0 & 0 & 0 & 0 & 0 & 0 & 0 & 0 & 0 & 0 & 0 & 0 & 0 & 0 & 0 \\
		0 & 0 & 0 & 0 & 0 & 0 & 0 & 0 & 0 & 0 & 0 & 0 & 0 & 0 & 0 & 0 & 1 & 0 & 0 & 0 & 0 & 0 & 0 & 0 & 0 & 0 & 0 & 0 & 1 & 0 & 0 & 0 \\
		0 & 0 & 0 & 0 & 0 & 0 & 0 & 0 & 0 & 0 & 0 & 0 & 0 & 0 & 0 & 0 & 0 & 1 & 0 & 0 & 0 & 0 & 0 & 0 & 0 & 0 & 0 & 0 & 0 & 0 & 0 & 0 \\
		0 & 0 & 0 & 0 & 0 & 0 & 0 & 0 & 0 & 0 & 0 & 0 & 0 & 0 & 0 & 0 & 0 & 0 & 1 & 0 & 1 & 0 & 0 & 0 & 0 & 0 & 0 & 0 & 0 & 0 & 0 & 0 \\
		0 & 0 & 0 & 0 & 0 & 0 & 0 & 0 & 0 & 0 & 0 & 0 & 0 & 0 & 0 & 0 & 0 & 0 & 0 & 1 & 0 & 0 & 0 & 0 & 0 & 0 & 0 & 0 & 0 & 0 & 0 & 0 \\
		0 & 0 & 0 & 0 & 0 & 0 & 0 & 0 & 0 & 0 & 0 & 0 & 0 & 0 & 0 & 0 & 0 & 0 & 1 & 0 & -1 & 0 & 0 & 0 & 0 & 0 & 0 & 0 & 0 & 0 & 0 & 0 \\
		0 & 0 & 0 & 0 & 0 & 0 & 0 & 0 & 0 & 0 & 0 & 0 & 0 & 0 & 0 & 0 & 0 & 0 & 0 & 0 & 0 & 1 & 0 & 0 & 0 & 0 & 0 & 0 & 0 & 0 & 0 & 0 \\
		0 & 0 & 0 & 0 & 0 & 0 & 0 & 0 & 0 & 0 & 0 & 0 & 0 & 0 & 0 & 0 & 0 & 0 & 0 & 0 & 0 & 0 & 1 & 0 & 1 & 0 & 0 & 0 & 0 & 0 & 0 & 0 \\
		0 & 0 & 0 & 0 & 0 & 0 & 0 & 0 & 0 & 0 & 0 & 0 & 0 & 0 & 0 & 0 & 0 & 0 & 0 & 0 & 0 & 0 & 0 & 1 & 0 & 0 & 0 & 0 & 0 & 0 & 0 & 0 \\
		0 & 0 & 0 & 0 & 0 & 0 & 0 & 0 & 0 & 0 & 0 & 0 & 0 & 0 & 0 & 0 & 0 & 0 & 0 & 0 & 0 & 0 & 1 & 0 & -1 & 0 & 0 & 0 & 0 & 0 & 0 & 0 \\
		0 & 0 & 0 & 0 & 0 & 0 & 0 & 0 & 0 & 0 & 0 & 0 & 0 & 0 & 0 & 0 & 0 & 0 & 0 & 0 & 0 & 0 & 0 & 0 & 0 & 1 & 0 & 0 & 0 & 0 & 0 & 0 \\
		0 & 0 & 0 & 0 & 0 & 0 & 0 & 0 & 0 & 0 & 0 & 0 & 0 & 0 & 0 & 0 & 0 & 0 & 0 & 0 & 0 & 0 & 0 & 0 & 0 & 0 & 1 & 0 & 0 & 0 & 2 & 0 \\
		0 & 0 & 0 & 0 & 0 & 0 & 0 & 0 & 0 & 0 & 0 & 0 & 0 & 0 & 0 & 0 & 0 & 0 & 0 & 0 & 0 & 0 & 0 & 0 & 0 & 0 & 0 & 1 & 0 & 0 & 0 & 0 \\
		0 & 0 & 0 & 0 & 0 & 0 & 0 & 0 & 0 & 0 & 0 & 0 & 0 & 0 & 0 & 0 & 1 & 0 & 0 & 0 & 0 & 0 & 0 & 0 & 0 & 0 & 0 & 0 & -1 & 0 & 0 & 0 \\
		0 & 0 & 0 & 0 & 0 & 0 & 0 & 0 & 0 & 0 & 0 & 0 & 0 & 0 & 0 & 0 & 0 & 0 & 0 & 0 & 0 & 0 & 0 & 0 & 0 & 0 & 0 & 0 & 0 & 1 & 0 & 0 \\
		0 & 0 & 0 & 0 & 0 & 0 & 0 & 0 & 0 & 0 & 0 & 0 & 0 & 0 & 0 & 0 & 0 & 0 & 0 & 0 & 0 & 0 & 0 & 0 & 0 & 0 & -2 & 0 & 0 & 0 & 1 & 0 \\
		0 & 0 & 0 & 0 & 0 & 0 & 0 & 0 & 0 & 0 & 0 & 0 & 0 & 0 & 0 & 0 & 0 & 0 & 0 & 0 & 0 & 0 & 0 & 0 & 0 & 0 & 0 & 0 & 0 & 0 & 0 & 1
	\end{bsmallmatrix*}
	,
\end{align}
\begin{align}
	\mathbf{W}_{11}
	=
	\begin{bsmallmatrix*}[r]
		\phantom{-}1 & \phantom{-}1 & \phantom{-}0 & \phantom{-}0 & \phantom{-}0 & \phantom{-}0 & \phantom{-}0 & \phantom{-}0 & \phantom{-}0 & \phantom{-}0 & \phantom{-}0 & \phantom{-}0 & \phantom{-}0 & \phantom{-}0 & \phantom{-}0 & \phantom{-}0 & \phantom{-}0 & \phantom{-}0 & \phantom{-}0 & \phantom{-}0 & \phantom{-}0 & \phantom{-}0 & \phantom{-}0 & \phantom{-}0 & \phantom{-}0 & \phantom{-}0 & \phantom{-}0 & \phantom{-}0 & \phantom{-}0 & \phantom{-}0 & \phantom{-}0 & \phantom{-}0 \\
		1 & -1 & 0 & 0 & 0 & 0 & 0 & 0 & 0 & 0 & 0 & 0 & 0 & 0 & 0 & 0 & 0 & 0 & 0 & 0 & 0 & 0 & 0 & 0 & 0 & 0 & 0 & 0 & 0 & 0 & 0 & 0 \\
		0 & 0 & 1 & 0 & 0 & 0 & 0 & 0 & 0 & 0 & 0 & 0 & 0 & 0 & 0 & 0 & 0 & 0 & 0 & 0 & 0 & 0 & 0 & 0 & 0 & 0 & 0 & 0 & 0 & 0 & 0 & 0 \\
		0 & 0 & 0 & 1 & 0 & 0 & 0 & 0 & 0 & 0 & 0 & 0 & 0 & 0 & 0 & 0 & 0 & 0 & 0 & 0 & 0 & 0 & 0 & 0 & 0 & 0 & 0 & 0 & 0 & 0 & 0 & 0 \\
		0 & 0 & 0 & 0 & 1 & 0 & 0 & 0 & 0 & 0 & 0 & 0 & 0 & 0 & 0 & 0 & 0 & 0 & 0 & 0 & 0 & 0 & 0 & 0 & 0 & 0 & 0 & 0 & 0 & 0 & 0 & 0 \\
		0 & 0 & 0 & 0 & 0 & 1 & 0 & 0 & 0 & 0 & 0 & 0 & 0 & 0 & 0 & 0 & 0 & 0 & 0 & 0 & 0 & 0 & 0 & 0 & 0 & 0 & 0 & 0 & 0 & 0 & 0 & 0 \\
		0 & 0 & 0 & 0 & 0 & 0 & 1 & 0 & 0 & 0 & 0 & 0 & 0 & 0 & 0 & 0 & 0 & 0 & 0 & 0 & 0 & 0 & 0 & 0 & 0 & 0 & 0 & 0 & 0 & 0 & 0 & 0 \\
		0 & 0 & 0 & 0 & 0 & 0 & 0 & 1 & 0 & 0 & 0 & 0 & 0 & 0 & 0 & 0 & 0 & 0 & 0 & 0 & 0 & 0 & 0 & 0 & 0 & 0 & 0 & 0 & 0 & 0 & 0 & 0 \\
		0 & 0 & 0 & 0 & 0 & 0 & 0 & 0 & 1 & 0 & 0 & 0 & 0 & 0 & 0 & 0 & 0 & 0 & 0 & 0 & 0 & 0 & 0 & 0 & 0 & 0 & 0 & 0 & 0 & 0 & 0 & 0 \\
		0 & 0 & 0 & 0 & 0 & 0 & 0 & 0 & 0 & 1 & 0 & 0 & 0 & 0 & 0 & 0 & 0 & 0 & 0 & 0 & 0 & 0 & 0 & 0 & 0 & 0 & 0 & 0 & 0 & 0 & 0 & 0 \\
		0 & 0 & 0 & 0 & 0 & 0 & 0 & 0 & 0 & 0 & 1 & 0 & 0 & 0 & 0 & 0 & 0 & 0 & 0 & 0 & 0 & 0 & 0 & 0 & 0 & 0 & 0 & 0 & 0 & 0 & 0 & 0 \\
		0 & 0 & 0 & 0 & 0 & 0 & 0 & 0 & 0 & 0 & 0 & 1 & 0 & 0 & 0 & 0 & 0 & 0 & 0 & 0 & 0 & 0 & 0 & 0 & 0 & 0 & 0 & 0 & 0 & 0 & 0 & 0 \\
		0 & 0 & 0 & 0 & 0 & 0 & 0 & 0 & 0 & 0 & 0 & 0 & 1 & 0 & 0 & 0 & 0 & 0 & 0 & 0 & 0 & 0 & 0 & 0 & 0 & 0 & 0 & 0 & 0 & 0 & 0 & 0 \\
		0 & 0 & 0 & 0 & 0 & 0 & 0 & 0 & 0 & 0 & 0 & 0 & 0 & 1 & 0 & 0 & 0 & 0 & 0 & 0 & 0 & 0 & 0 & 0 & 0 & 0 & 0 & 0 & 0 & 0 & 0 & 0 \\
		0 & 0 & 0 & 0 & 0 & 0 & 0 & 0 & 0 & 0 & 0 & 0 & 0 & 0 & 1 & 0 & 0 & 0 & 0 & 0 & 0 & 0 & 0 & 0 & 0 & 0 & 0 & 0 & 0 & 0 & 0 & 0 \\
		0 & 0 & 0 & 0 & 0 & 0 & 0 & 0 & 0 & 0 & 0 & 0 & 0 & 0 & 0 & 1 & 0 & 0 & 0 & 0 & 0 & 0 & 0 & 0 & 0 & 0 & 0 & 0 & 0 & 0 & 0 & 0 \\
		0 & 0 & 0 & 0 & 0 & 0 & 0 & 0 & 0 & 0 & 0 & 0 & 0 & 0 & 0 & 0 & 1 & 0 & 0 & 0 & 0 & 0 & 0 & 0 & 0 & 0 & 1 & 0 & 0 & 0 & 0 & 0 \\
		0 & 0 & 0 & 0 & 0 & 0 & 0 & 0 & 0 & 0 & 0 & 0 & 0 & 0 & 0 & 0 & 0 & 1 & 1 & 0 & 0 & 0 & 0 & 0 & 0 & 0 & 0 & 0 & 0 & 0 & 0 & 0 \\
		0 & 0 & 0 & 0 & 0 & 0 & 0 & 0 & 0 & 0 & 0 & 0 & 0 & 0 & 0 & 0 & 0 & 1 & -1 & 0 & 0 & 0 & 0 & 0 & 0 & 0 & 0 & 0 & 0 & 0 & 0 & 0 \\
		0 & 0 & 0 & 0 & 0 & 0 & 0 & 0 & 0 & 0 & 0 & 0 & 0 & 0 & 0 & 0 & 0 & 0 & 0 & 1 & 0 & 0 & 1 & 0 & 0 & 0 & 0 & 0 & 0 & 0 & 0 & 0 \\
		0 & 0 & 0 & 0 & 0 & 0 & 0 & 0 & 0 & 0 & 0 & 0 & 0 & 0 & 0 & 0 & 0 & 0 & 0 & 0 & 1 & 0 & 0 & 1 & 0 & 0 & 0 & 0 & 0 & 0 & 0 & 0 \\
		0 & 0 & 0 & 0 & 0 & 0 & 0 & 0 & 0 & 0 & 0 & 0 & 0 & 0 & 0 & 0 & 0 & 0 & 0 & 0 & 0 & 1 & 0 & 0 & 1 & 0 & 0 & 0 & 0 & 0 & 0 & 0 \\
		0 & 0 & 0 & 0 & 0 & 0 & 0 & 0 & 0 & 0 & 0 & 0 & 0 & 0 & 0 & 0 & 0 & 0 & 0 & 1 & 0 & 0 & -1 & 0 & 0 & 0 & 0 & 0 & 0 & 0 & 0 & 0 \\
		0 & 0 & 0 & 0 & 0 & 0 & 0 & 0 & 0 & 0 & 0 & 0 & 0 & 0 & 0 & 0 & 0 & 0 & 0 & 0 & 1 & 0 & 0 & -1 & 0 & 0 & 0 & 0 & 0 & 0 & 0 & 0 \\
		0 & 0 & 0 & 0 & 0 & 0 & 0 & 0 & 0 & 0 & 0 & 0 & 0 & 0 & 0 & 0 & 0 & 0 & 0 & 0 & 0 & 1 & 0 & 0 & -1 & 0 & 0 & 0 & 0 & 0 & 0 & 0 \\
		0 & 0 & 0 & 0 & 0 & 0 & 0 & 0 & 0 & 0 & 0 & 0 & 0 & 0 & 0 & 0 & 0 & 0 & 0 & 0 & 0 & 0 & 0 & 0 & 0 & 1 & 0 & 0 & 0 & 0 & 0 & 0 \\
		0 & 0 & 0 & 0 & 0 & 0 & 0 & 0 & 0 & 0 & 0 & 0 & 0 & 0 & 0 & 0 & 1 & 0 & 0 & 0 & 0 & 0 & 0 & 0 & 0 & 0 & -1 & 0 & 0 & 0 & 0 & 0 \\
		0 & 0 & 0 & 0 & 0 & 0 & 0 & 0 & 0 & 0 & 0 & 0 & 0 & 0 & 0 & 0 & 0 & 0 & 0 & 0 & 0 & 0 & 0 & 0 & 0 & 0 & 0 & 1 & 0 & 0 & 0 & 0 \\
		0 & 0 & 0 & 0 & 0 & 0 & 0 & 0 & 0 & 0 & 0 & 0 & 0 & 0 & 0 & 0 & 0 & 0 & 0 & 0 & 0 & 0 & 0 & 0 & 0 & 0 & 0 & 0 & 1 & 0 & 1 & 0 \\
		0 & 0 & 0 & 0 & 0 & 0 & 0 & 0 & 0 & 0 & 0 & 0 & 0 & 0 & 0 & 0 & 0 & 0 & 0 & 0 & 0 & 0 & 0 & 0 & 0 & 0 & 0 & 0 & 0 & 1 & 0 & 0 \\
		0 & 0 & 0 & 0 & 0 & 0 & 0 & 0 & 0 & 0 & 0 & 0 & 0 & 0 & 0 & 0 & 0 & 0 & 0 & 0 & 0 & 0 & 0 & 0 & 0 & 0 & 0 & 0 & 1 & 0 & -1 & 0 \\
		0 & 0 & 0 & 0 & 0 & 0 & 0 & 0 & 0 & 0 & 0 & 0 & 0 & 0 & 0 & 0 & 0 & 0 & 0 & 0 & 0 & 0 & 0 & 0 & 0 & 0 & 0 & 0 & 0 & 0 & 0 & 1
	\end{bsmallmatrix*}
	,
\end{align}
\begin{align}
	\mathbf{W}_{12}
	=
	\begin{bsmallmatrix*}[r]
		\phantom{-}1 & \phantom{-}0 & \phantom{-}0 & \phantom{-}0 & \phantom{-}0 & \phantom{-}0 & \phantom{-}0 & \phantom{-}0 & \phantom{-}0 & \phantom{-}0 & \phantom{-}0 & \phantom{-}0 & \phantom{-}0 & \phantom{-}0 & \phantom{-}0 & \phantom{-}0 & \phantom{-}0 & \phantom{-}0 & \phantom{-}0 & \phantom{-}0 & \phantom{-}0 & \phantom{-}0 & \phantom{-}0 & \phantom{-}0 & \phantom{-}0 & \phantom{-}0 & \phantom{-}0 & \phantom{-}0 & \phantom{-}0 & \phantom{-}0 & \phantom{-}0 & \phantom{-}0 \\
		0 & 1 & 0 & 0 & 0 & 0 & 0 & 0 & 0 & 0 & 0 & 0 & 0 & 0 & 0 & 0 & 0 & 0 & 0 & 0 & 0 & 0 & 0 & 0 & 0 & 0 & 0 & 0 & 0 & 0 & 0 & 0 \\
		0 & 0 & 1 & 0 & 0 & 0 & 0 & 0 & 0 & 0 & 0 & 0 & 0 & 0 & 0 & 0 & 0 & 0 & 0 & 0 & 0 & 0 & 0 & 0 & 0 & 0 & 0 & 0 & 0 & 0 & 0 & 0 \\
		0 & 0 & 0 & 1 & 0 & 0 & 0 & 0 & 0 & 0 & 0 & 0 & 0 & 0 & 0 & 0 & 0 & 0 & 0 & 0 & 0 & 0 & 0 & 0 & 0 & 0 & 0 & 0 & 0 & 0 & 0 & 0 \\
		0 & 0 & 0 & 0 & 1 & 0 & 0 & 0 & 0 & 0 & 0 & 0 & 0 & 0 & 0 & 0 & 0 & 0 & 0 & 0 & 0 & 0 & 0 & 0 & 0 & 0 & 0 & 0 & 0 & 0 & 0 & 0 \\
		0 & 0 & 0 & 0 & 0 & 1 & 0 & 0 & 0 & 0 & 0 & 0 & 0 & 0 & 0 & 0 & 0 & 0 & 0 & 0 & 0 & 0 & 0 & 0 & 0 & 0 & 0 & 0 & 0 & 0 & 0 & 0 \\
		0 & 0 & 0 & 0 & 0 & 0 & 1 & 0 & 0 & 0 & 0 & 0 & 0 & 0 & 0 & 0 & 0 & 0 & 0 & 0 & 0 & 0 & 0 & 0 & 0 & 0 & 0 & 0 & 0 & 0 & 0 & 0 \\
		0 & 0 & 0 & 0 & 0 & 0 & 0 & 1 & 0 & 0 & 0 & 0 & 0 & 0 & 0 & 0 & 0 & 0 & 0 & 0 & 0 & 0 & 0 & 0 & 0 & 0 & 0 & 0 & 0 & 0 & 0 & 0 \\
		0 & 0 & 0 & 0 & 0 & 0 & 0 & 0 & 1 & 0 & 0 & 0 & 0 & 0 & 0 & 0 & 0 & 0 & 0 & 0 & 0 & 0 & 0 & 0 & 0 & 0 & 0 & 0 & 0 & 0 & 0 & 0 \\
		0 & 0 & 0 & 0 & 0 & 0 & 0 & 0 & 0 & 1 & 0 & 0 & 0 & 0 & 0 & 0 & 0 & 0 & 0 & 0 & 0 & 0 & 0 & 0 & 0 & 0 & 0 & 0 & 0 & 0 & 0 & 0 \\
		0 & 0 & 0 & 0 & 0 & 0 & 0 & 0 & 0 & 0 & 1 & 0 & 0 & 0 & 0 & 0 & 0 & 0 & 0 & 0 & 0 & 0 & 0 & 0 & 0 & 0 & 0 & 0 & 0 & 0 & 0 & 0 \\
		0 & 0 & 0 & 0 & 0 & 0 & 0 & 0 & 0 & 0 & 0 & 1 & 0 & 0 & 0 & 0 & 0 & 0 & 0 & 0 & 0 & 0 & 0 & 0 & 0 & 0 & 0 & 0 & 0 & 0 & 0 & 0 \\
		0 & 0 & 0 & 0 & 0 & 0 & 0 & 0 & 0 & 0 & 0 & 0 & 1 & 0 & 0 & 0 & 0 & 0 & 0 & 0 & 0 & 0 & 0 & 0 & 0 & 0 & 0 & 0 & 0 & 0 & 0 & 0 \\
		0 & 0 & 0 & 0 & 0 & 0 & 0 & 0 & 0 & 0 & 0 & 0 & 0 & 1 & 0 & 0 & 0 & 0 & 0 & 0 & 0 & 0 & 0 & 0 & 0 & 0 & 0 & 0 & 0 & 0 & 0 & 0 \\
		0 & 0 & 0 & 0 & 0 & 0 & 0 & 0 & 0 & 0 & 0 & 0 & 0 & 0 & 1 & 0 & 0 & 0 & 0 & 0 & 0 & 0 & 0 & 0 & 0 & 0 & 0 & 0 & 0 & 0 & 0 & 0 \\
		0 & 0 & 0 & 0 & 0 & 0 & 0 & 0 & 0 & 0 & 0 & 0 & 0 & 0 & 0 & 1 & 0 & 0 & 0 & 0 & 0 & 0 & 0 & 0 & 0 & 0 & 0 & 0 & 0 & 0 & 0 & 0 \\
		0 & 0 & 0 & 0 & 0 & 0 & 0 & 0 & 0 & 0 & 0 & 0 & 0 & 0 & 0 & 0 & 1 & 0 & 0 & 0 & 0 & 0 & 0 & 0 & 0 & 1 & 0 & 0 & 0 & 0 & 0 & 0 \\
		0 & 0 & 0 & 0 & 0 & 0 & 0 & 0 & 0 & 0 & 0 & 0 & 0 & 0 & 0 & 0 & 0 & 1 & 0 & 0 & 0 & 0 & 0 & 0 & 0 & 0 & 0 & 0 & 0 & 0 & 0 & 0 \\
		0 & 0 & 0 & 0 & 0 & 0 & 0 & 0 & 0 & 0 & 0 & 0 & 0 & 0 & 0 & 0 & 0 & 0 & 1 & 0 & 0 & 0 & 0 & 0 & 0 & 0 & 0 & 0 & 0 & 0 & 0 & 0 \\
		0 & 0 & 0 & 0 & 0 & 0 & 0 & 0 & 0 & 0 & 0 & 0 & 0 & 0 & 0 & 0 & 0 & 0 & 0 & 1 & 0 & 0 & 0 & 0 & 0 & 0 & 0 & 0 & 0 & 0 & 0 & 0 \\
		0 & 0 & 0 & 0 & 0 & 0 & 0 & 0 & 0 & 0 & 0 & 0 & 0 & 0 & 0 & 0 & 0 & 0 & 0 & 0 & 1 & 0 & 0 & 0 & 0 & 0 & 0 & 0 & 0 & 0 & 0 & 0 \\
		0 & 0 & 0 & 0 & 0 & 0 & 0 & 0 & 0 & 0 & 0 & 0 & 0 & 0 & 0 & 0 & 0 & 0 & 0 & 0 & 0 & 1 & 0 & 0 & 0 & 0 & 0 & 0 & 0 & 0 & 0 & 0 \\
		0 & 0 & 0 & 0 & 0 & 0 & 0 & 0 & 0 & 0 & 0 & 0 & 0 & 0 & 0 & 0 & 0 & 0 & 0 & 0 & 0 & 0 & 1 & 0 & 0 & 0 & 0 & 0 & 0 & 0 & 0 & 0 \\
		0 & 0 & 0 & 0 & 0 & 0 & 0 & 0 & 0 & 0 & 0 & 0 & 0 & 0 & 0 & 0 & 0 & 0 & 0 & 0 & 0 & 0 & 0 & 1 & 0 & 0 & 0 & 0 & 0 & 0 & 0 & 0 \\
		0 & 0 & 0 & 0 & 0 & 0 & 0 & 0 & 0 & 0 & 0 & 0 & 0 & 0 & 0 & 0 & 0 & 0 & 0 & 0 & 0 & 0 & 0 & 0 & 1 & 0 & 0 & 0 & 0 & 0 & 0 & 0 \\
		0 & 0 & 0 & 0 & 0 & 0 & 0 & 0 & 0 & 0 & 0 & 0 & 0 & 0 & 0 & 0 & 1 & 0 & 0 & 0 & 0 & 0 & 0 & 0 & 0 & -1 & 0 & 0 & 0 & 0 & 0 & 0 \\
		0 & 0 & 0 & 0 & 0 & 0 & 0 & 0 & 0 & 0 & 0 & 0 & 0 & 0 & 0 & 0 & 0 & 0 & 0 & 0 & 0 & 0 & 0 & 0 & 0 & 0 & 1 & 0 & 0 & 0 & 0 & 1 \\
		0 & 0 & 0 & 0 & 0 & 0 & 0 & 0 & 0 & 0 & 0 & 0 & 0 & 0 & 0 & 0 & 0 & 0 & 0 & 0 & 0 & 0 & 0 & 0 & 0 & 0 & 0 & 1 & 1 & 0 & 0 & 0 \\
		0 & 0 & 0 & 0 & 0 & 0 & 0 & 0 & 0 & 0 & 0 & 0 & 0 & 0 & 0 & 0 & 0 & 0 & 0 & 0 & 0 & 0 & 0 & 0 & 0 & 0 & 0 & 1 & -1 & 0 & 0 & 0 \\
		0 & 0 & 0 & 0 & 0 & 0 & 0 & 0 & 0 & 0 & 0 & 0 & 0 & 0 & 0 & 0 & 0 & 0 & 0 & 0 & 0 & 0 & 0 & 0 & 0 & 0 & 0 & 0 & 0 & 1 & 1 & 0 \\
		0 & 0 & 0 & 0 & 0 & 0 & 0 & 0 & 0 & 0 & 0 & 0 & 0 & 0 & 0 & 0 & 0 & 0 & 0 & 0 & 0 & 0 & 0 & 0 & 0 & 0 & 0 & 0 & 0 & 1 & -1 & 0 \\
		0 & 0 & 0 & 0 & 0 & 0 & 0 & 0 & 0 & 0 & 0 & 0 & 0 & 0 & 0 & 0 & 0 & 0 & 0 & 0 & 0 & 0 & 0 & 0 & 0 & 0 & 1 & 0 & 0 & 0 & 0 & -1
	\end{bsmallmatrix*}
	,
\end{align}
and
\begin{align}
	\mathbf{W}_{13}
	=
	\begin{bsmallmatrix*}[r]
		\phantom{-}2 & \phantom{-}0 & \phantom{-}0 & \phantom{-}0 & \phantom{-}0 & \phantom{-}0 & \phantom{-}0 & \phantom{-}0 & \phantom{-}0 & \phantom{-}0 & \phantom{-}0 & \phantom{-}0 & \phantom{-}0 & \phantom{-}0 & \phantom{-}0 & \phantom{-}0 & \phantom{-}0 & \phantom{-}0 & \phantom{-}0 & \phantom{-}0 & \phantom{-}0 & \phantom{-}0 & \phantom{-}0 & \phantom{-}0 & \phantom{-}0 & \phantom{-}0 & \phantom{-}0 & \phantom{-}0 & \phantom{-}0 & \phantom{-}0 & \phantom{-}0 & \phantom{-}0 \\
		0 & 0 & 0 & 0 & 0 & 0 & 0 & 0 & 0 & 0 & 0 & 0 & 0 & 0 & 0 & 0 & 0 & 0 & j & 0 & 0 & 0 & 0 & 0 & 0 & 1 & 0 & 0 & 0 & 0 & 0 & 0 \\
		0 & 0 & 0 & 0 & 0 & 1 & 0 & 0 & 0 & -j & 0 & 0 & 0 & 0 & 0 & 0 & 0 & 0 & 0 & 0 & 0 & 0 & 0 & 0 & 0 & 0 & 0 & 0 & 0 & 0 & 0 & 0 \\
		0 & 0 & 0 & 0 & 0 & 0 & 0 & 0 & 0 & 0 & 0 & 0 & 0 & 0 & 0 & 0 & 0 & 0 & 0 & j & 0 & 0 & 0 & 0 & 0 & 0 & 0 & 1 & 0 & 0 & 0 & 0 \\
		0 & 0 & 0 & 1 & 0 & 0 & 0 & 0 & 0 & 0 & 0 & 0 & 0 & j & 0 & 0 & 0 & 0 & 0 & 0 & 0 & 0 & 0 & 0 & 0 & 0 & 0 & 0 & 0 & 0 & 0 & 0 \\
		0 & 0 & 0 & 0 & 0 & 0 & 0 & 0 & 0 & 0 & 0 & 0 & 0 & 0 & 0 & 0 & 0 & 0 & 0 & 0 & 0 & j & 0 & 0 & 0 & 0 & 0 & 0 & 0 & 0 & -1 & 0 \\
		0 & 0 & 0 & 0 & 0 & 0 & 0 & 1 & 0 & 0 & 0 & j & 0 & 0 & 0 & 0 & 0 & 0 & 0 & 0 & 0 & 0 & 0 & 0 & 0 & 0 & 0 & 0 & 0 & 0 & 0 & 0 \\
		0 & 0 & 0 & 0 & 0 & 0 & 0 & 0 & 0 & 0 & 0 & 0 & 0 & 0 & 0 & 0 & 0 & 0 & 0 & 0 & 0 & 0 & 0 & -j & 0 & 0 & 1 & 0 & 0 & 0 & 0 & 0 \\
		0 & 0 & 1 & 0 & 0 & 0 & 0 & 0 & 0 & 0 & 0 & 0 & 0 & 0 & 0 & -j & 0 & 0 & 0 & 0 & 0 & 0 & 0 & 0 & 0 & 0 & 0 & 0 & 0 & 0 & 0 & 0 \\
		0 & 0 & 0 & 0 & 0 & 0 & 0 & 0 & 0 & 0 & 0 & 0 & 0 & 0 & 0 & 0 & 0 & 0 & 0 & 0 & j & 0 & 0 & 0 & 0 & 0 & 0 & 0 & 0 & 0 & 0 & 1 \\
		0 & 0 & 0 & 0 & 0 & 0 & 0 & 0 & -1 & 0 & 0 & 0 & j & 0 & 0 & 0 & 0 & 0 & 0 & 0 & 0 & 0 & 0 & 0 & 0 & 0 & 0 & 0 & 0 & 0 & 0 & 0 \\
		0 & 0 & 0 & 0 & 0 & 0 & 0 & 0 & 0 & 0 & 0 & 0 & 0 & 0 & 0 & 0 & 0 & 0 & 0 & 0 & 0 & 0 & 0 & 0 & j & 0 & 0 & 0 & 0 & 1 & 0 & 0 \\
		0 & 0 & 0 & 0 & 1 & 0 & 0 & 0 & 0 & 0 & 0 & 0 & 0 & 0 & -j & 0 & 0 & 0 & 0 & 0 & 0 & 0 & 0 & 0 & 0 & 0 & 0 & 0 & 0 & 0 & 0 & 0 \\
		0 & 0 & 0 & 0 & 0 & 0 & 0 & 0 & 0 & 0 & 0 & 0 & 0 & 0 & 0 & 0 & 0 & 0 & 0 & 0 & 0 & 0 & j & 0 & 0 & 0 & 0 & 0 & -1 & 0 & 0 & 0 \\
		0 & 0 & 0 & 0 & 0 & 0 & -1 & 0 & 0 & 0 & -j & 0 & 0 & 0 & 0 & 0 & 0 & 0 & 0 & 0 & 0 & 0 & 0 & 0 & 0 & 0 & 0 & 0 & 0 & 0 & 0 & 0 \\
		0 & 0 & 0 & 0 & 0 & 0 & 0 & 0 & 0 & 0 & 0 & 0 & 0 & 0 & 0 & 0 & 1 & j & 0 & 0 & 0 & 0 & 0 & 0 & 0 & 0 & 0 & 0 & 0 & 0 & 0 & 0 \\
		0 & 2 & 0 & 0 & 0 & 0 & 0 & 0 & 0 & 0 & 0 & 0 & 0 & 0 & 0 & 0 & 0 & 0 & 0 & 0 & 0 & 0 & 0 & 0 & 0 & 0 & 0 & 0 & 0 & 0 & 0 & 0 \\
		0 & 0 & 0 & 0 & 0 & 0 & 0 & 0 & 0 & 0 & 0 & 0 & 0 & 0 & 0 & 0 & 1 & -j & 0 & 0 & 0 & 0 & 0 & 0 & 0 & 0 & 0 & 0 & 0 & 0 & 0 & 0 \\
		0 & 0 & 0 & 0 & 0 & 0 & -1 & 0 & 0 & 0 & j & 0 & 0 & 0 & 0 & 0 & 0 & 0 & 0 & 0 & 0 & 0 & 0 & 0 & 0 & 0 & 0 & 0 & 0 & 0 & 0 & 0 \\
		0 & 0 & 0 & 0 & 0 & 0 & 0 & 0 & 0 & 0 & 0 & 0 & 0 & 0 & 0 & 0 & 0 & 0 & 0 & 0 & 0 & 0 & -j & 0 & 0 & 0 & 0 & 0 & -1 & 0 & 0 & 0 \\
		0 & 0 & 0 & 0 & 1 & 0 & 0 & 0 & 0 & 0 & 0 & 0 & 0 & 0 & j & 0 & 0 & 0 & 0 & 0 & 0 & 0 & 0 & 0 & 0 & 0 & 0 & 0 & 0 & 0 & 0 & 0 \\
		0 & 0 & 0 & 0 & 0 & 0 & 0 & 0 & 0 & 0 & 0 & 0 & 0 & 0 & 0 & 0 & 0 & 0 & 0 & 0 & 0 & 0 & 0 & 0 & -j & 0 & 0 & 0 & 0 & 1 & 0 & 0 \\
		0 & 0 & 0 & 0 & 0 & 0 & 0 & 0 & -1 & 0 & 0 & 0 & -j & 0 & 0 & 0 & 0 & 0 & 0 & 0 & 0 & 0 & 0 & 0 & 0 & 0 & 0 & 0 & 0 & 0 & 0 & 0 \\
		0 & 0 & 0 & 0 & 0 & 0 & 0 & 0 & 0 & 0 & 0 & 0 & 0 & 0 & 0 & 0 & 0 & 0 & 0 & 0 & -j & 0 & 0 & 0 & 0 & 0 & 0 & 0 & 0 & 0 & 0 & 1 \\
		0 & 0 & 1 & 0 & 0 & 0 & 0 & 0 & 0 & 0 & 0 & 0 & 0 & 0 & 0 & j & 0 & 0 & 0 & 0 & 0 & 0 & 0 & 0 & 0 & 0 & 0 & 0 & 0 & 0 & 0 & 0 \\
		0 & 0 & 0 & 0 & 0 & 0 & 0 & 0 & 0 & 0 & 0 & 0 & 0 & 0 & 0 & 0 & 0 & 0 & 0 & 0 & 0 & 0 & 0 & j & 0 & 0 & 1 & 0 & 0 & 0 & 0 & 0 \\
		0 & 0 & 0 & 0 & 0 & 0 & 0 & 1 & 0 & 0 & 0 & -j & 0 & 0 & 0 & 0 & 0 & 0 & 0 & 0 & 0 & 0 & 0 & 0 & 0 & 0 & 0 & 0 & 0 & 0 & 0 & 0 \\
		0 & 0 & 0 & 0 & 0 & 0 & 0 & 0 & 0 & 0 & 0 & 0 & 0 & 0 & 0 & 0 & 0 & 0 & 0 & 0 & 0 & -j & 0 & 0 & 0 & 0 & 0 & 0 & 0 & 0 & -1 & 0 \\
		0 & 0 & 0 & 1 & 0 & 0 & 0 & 0 & 0 & 0 & 0 & 0 & 0 & -j & 0 & 0 & 0 & 0 & 0 & 0 & 0 & 0 & 0 & 0 & 0 & 0 & 0 & 0 & 0 & 0 & 0 & 0 \\
		0 & 0 & 0 & 0 & 0 & 0 & 0 & 0 & 0 & 0 & 0 & 0 & 0 & 0 & 0 & 0 & 0 & 0 & 0 & -j & 0 & 0 & 0 & 0 & 0 & 0 & 0 & 1 & 0 & 0 & 0 & 0 \\
		0 & 0 & 0 & 0 & 0 & 1 & 0 & 0 & 0 & j & 0 & 0 & 0 & 0 & 0 & 0 & 0 & 0 & 0 & 0 & 0 & 0 & 0 & 0 & 0 & 0 & 0 & 0 & 0 & 0 & 0 & 0 \\
		0 & 0 & 0 & 0 & 0 & 0 & 0 & 0 & 0 & 0 & 0 & 0 & 0 & 0 & 0 & 0 & 0 & 0 & -j & 0 & 0 & 0 & 0 & 0 & 0 & 1 & 0 & 0 & 0 & 0 & 0 & 0
	\end{bsmallmatrix*}
	.
\end{align}

{\small
\singlespacing
\bibliographystyle{ieeetr}
\bibliography{bib}

@book{kay1998fundamentals,
    author = {S. M. Kay},
    title = {Fundamentals of Statistical Signal Processing: Detection Theory},
    publisher = {Prentice-Hall PTR},
    year = {1998},
    address = {Upper Saddle River, NJ},
    volume = {II},
}

@Book{burton2011elementary,
  author    = {D. M. Burton},
  title     = {Elementary Number Theory},
  publisher = {McGraw-Hill},
  year      = {2011},
  address   = {New York, NY},
  edition   = {7th}
}

@MastersThesis{suarezvillagran2015aproximacoes,
  author    = {D. M. {Su\'arez Villagr\'an}},
  title     = {Discrete {F}ourier transform approximations with applications in detection and estimation},
  school    = {Universidade Federal de Pernambuco},
  year      = {2015},
  address   = {Recife, Brazil},
  note      = {Advisors: Dr~R~J~Cintra and Dr~F~M~Bayer},
}

@Article{portella2022radix,
  author    = {L. Portella and D. F. G. Coelho and F. M. Bayer and A. Madanayake and R. J. Cintra},
  title     = {Radix-${N}$ Algorithm for Computing ${N}^{2^n}$-point {DFT} Approximations},
  journal   = {{IEEE} Signal Processing Letters},
  year      = {2022},
  volume    = {29},
  pages     = {1838--1842},
  doi       = {10.1109/lsp.2022.3200573},
  url       = {https://ieeexplore.ieee.org/document/9864023}
}

@book{oppenheim1999discrete,
  title={Discrete-time Signal Processing},
  author={Oppenheim, A. V. and Schafer, R. W. and Buck, J.R.},
  isbn={9780137549207},
  lccn={98050398},
  year={1999},
  address = {Upper Saddle River, NJ},
  publisher={Prentice Hall}
}

@Article{cintra2024approximation,
  author      = {R. J. Cintra},
  title       = {An Approximation for the 32-point Discrete {F}ourier Transform},
  year        = {2024},
  date        = {2024-07-17},
  doi         = {10.48550/arXiv.2407.12708},
  eprint      = {2407.12708v2},
  eprintclass = {eess.SP},
  journal     = {arXiv},
  url         = {https://arxiv.org/abs/2407.12708},
  note        = {10.48550/arXiv.2407.12708},
}

@Article{cintra2011integer,
  author    = {R. J. Cintra},
  title     = {An Integer Approximation Method for Discrete Sinusoidal Transforms},
  journal   = {Circuits, Systems, and Signal Processing},
  year      = {2011},
  volume    = {30},
  number    = {6},
  pages     = {1481--1501},
  doi       = {10.1007/s00034-011-9318-5},
  eprint    = {2007.02232},
  publisher = {Springer Science and Business Media {LLC}},
  url       = {https://link.springer.com/article/10.1007/s00034-011-9318-5},
}

@Article{coelho2024discrete,
  author      = {D. F. G. Coelho and R. J. Cintra},
  title       = {Discrete {F}ourier Transform Approximations Based on the {C}ooley-{T}ukey Radix-2 Algorithm},
  journal     = {arXiv},
  year        = {2024},
  date        = {2024-02-25},
  doi         = {10.48550/arXiv.2402.16225},
  eprint      = {2402.16225},
  eprintclass = {eess.SP},
  url         = {https://arxiv.org/abs/2402.16225},
  note        = {10.48550/arXiv.2402.16225},
}

@Article{portella2025multiplierless,
  author    = {L. Portella and F. M. Bayer and R. J. Cintra},
  title     = {Multiplierless {DFT} Approximation Based on the Prime Factor Algorithm},
  journal   = {IEEE Transactions on Signal Processing},
  year      = {2025},
  volume    = {73},
  pages     = {5273--5285},
  doi       = {10.1109/tsp.2025.3634427},
  url       = {https://ieeexplore.ieee.org/document/11261682},
}

@book{blahut2010,
  author    = {Blahut, R. E.},
  title     = {Fast Algorithms for Digital Signal Processing},
  publisher = {CUP},
  address   = {Cambridge, UK},
  year      = {2010}
}

@book{seber2008matrix,
  author = {Seber, G. A.},
  title = {A Matrix Handbook for Statisticians},
  publisher = {Wiley},
  address = {Hoboken, NJ},
  year = {2008}
}

@Article{cintra2011dct,
  author   = {Cintra, R. J. and Bayer, F. M.},
  title    = {A {DCT} approximation for image compression},
  journal  = {IEEE Signal Process Letters},
  year     = {2011},
  volume   = {18},
  number   = {10},
  pages    = {579--582},
  doi      = {10.1109/LSP.2011.2163394},
  fjournal = {IEEE Signal Processing Letters},
}

@Book{Hyndman2018,
  author    = {Hyndman, R. J. and Athanasopoulos, G.},
  title     = {Forecasting: Principles and Practice},
  edition   = {2nd},
  publisher = {OTexts},
  year      = {2018},
  address   = {Melbourne, Australia},
  isbn      = {978-0-9875071-1-2},
}

@book{herstein1975,
  author    = {Herstein, I. N.},
  title     = {Topics in Algebra},
  publisher = {John Wiley \& Sons},
  address   = {New York, NY},
  year      = {1975}
}

@book{cormen2009introduction,
  author    = {T. H. Cormen and C. E. Leiserson
               and Ronald L. Rivest and Clifford Stein},
  title     = {Introduction to Algorithms},
  edition   = {3rd},
  publisher = {MIT Press},
  address   = {Cambridge, MA},
  year      = {2009},
  isbn      = {978-0-262-03384-8}
}

@book{rosen2019discrete,
  author    = {K. H. Rosen},
  title     = {Discrete Mathematics and Its Applications},
  edition   = {8th},
  publisher = {McGraw-Hill Education},
  address   = {New York, NY},
  year      = {2019},
  isbn      = {9781260091991}
}

@article{weinstein1971data,
  author  = {Weinstein, S. and Ebert, P.},
  title   = {Data Transmission by Frequency-Division Multiplexing
             Using the Discrete {F}ourier Transform},
  journal = {IEEE Transactions on Communication Technology},
  volume  = {19},
  number  = {5},
  pages   = {628--634},
  year    = {1971},
  doi     = {10.1109/TCOM.1971.1090705}
}

@article{jaradat2019modulation,
  author  = {Jaradat, A. M. and Hamamreh, J. M. and Arslan, H.},
  title   = {Modulation Options for {OFDM}-Based Waveforms:
             Classification, Comparison, and Future Directions},
  journal = {IEEE Access},
  volume  = {7},
  pages   = {17263--17278},
  year    = {2019},
  doi     = {10.1109/ACCESS.2019.2895958}
}

@article{ylikaakinen2021frequency,
  author  = {Yli-Kaakinen, J. and Loulou, A. and Levanen, T. and
             Pajukoski, K. and Palin, A. and Renfors, M. and Valkama, M.},
  title   = {Frequency-Domain Signal Processing for Spectrally-Enhanced
             {CP-OFDM} Waveforms in {5G} New Radio},
  journal = {IEEE Transactions on Wireless Communications},
  volume  = {20},
  number  = {10},
  pages   = {6867--6883},
  year    = {2021},
  doi     = {10.1109/TWC.2021.3077762}
}

@article{wang2024low,
  author  = {Wang, L. and Ai, Z. and Shi, J. and Wang, J. and
             Zhang, X. and Yang, J.},
  title   = {Low Computational Complexity {SAR} Imaging Algorithm
             for Ship Monitoring via 2-{D} Band-Limited Sparse
             {F}ourier Transform},
  journal = {IEEE Sensors Journal},
  volume  = {24},
  number  = {8},
  pages   = {13326--13342},
  year    = {2024},
  doi     = {10.1109/JSEN.2024.3370234}
}

@article{flury1986algorithm,
  author  = {Flury, B. N. and Gautschi, W.},
  title   = {An Algorithm for Simultaneous Orthogonal Transformation of Several Positive Definite Symmetric Matrices to Nearly Diagonal Form},
  journal = {SIAM Journal on Scientific and Statistical Computing},
  volume  = {7},
  number  = {1},
  pages   = {169--184},
  year    = {1986},
  doi     = {10.1137/0907013},
}

@article{agrez2005improving,
  author  = {Agrez, D.},
  title   = {Improving Phase Estimation With Leakage Minimization},
  journal = {IEEE Transactions on Instrumentation and Measurement},
  volume  = {54},
  number  = {4},
  pages   = {1347--1353},
  year    = {2005},
  doi     = {10.1109/TIM.2005.851058},
}

@article{quinn1997estimation,
  author  = {Quinn, B. G.},
  title   = {Estimation of Frequency, Amplitude, and Phase From the
             {DFT} of a Time Series},
  journal = {IEEE Transactions on Signal Processing},
  volume  = {45},
  number  = {3},
  pages   = {814--817},
  year    = {1997},
  doi     = {10.1109/78.558515},
}
}

\end{document}